\PassOptionsToPackage{hypertexnames=false}{hyperref}
\documentclass[journal]{IEEEtran}
\usepackage{pifont}
\newcommand{\cmark}{\ding{51}}  % checkmark
\newcommand{\xmark}{\ding{55}}  % x-mark
\usepackage{mathtools}
\usepackage{xcolor}
\usepackage{amsmath}
\usepackage{cite}
\usepackage{makecell}
\usepackage{hyperref}
\usepackage{array,booktabs,multirow}
\usepackage{soul}

 \usepackage{placeins} % Add this in the preamble for \FloatBarrier
\usepackage{algorithm}
\usepackage{algpseudocode}
\usepackage[pdftex]{graphicx}
\usepackage{amssymb}
\usepackage{MnSymbol}
\usepackage{float} % Add this line
\usepackage{dblfloatfix} % Allow double-column figures at the bottom of a page
\usepackage[caption=false,font=footnotesize]{subfig}

\usepackage{varwidth}

\begin{document}
\bstctlcite{IEEEtran:BSTcontrol}

\markboth{Rai \MakeLowercase{\textit{et al.}}: Multi-Target Micro-Motion Parameter Estimation}%
{Rai \MakeLowercase{\textit{et al.}}: Multi-Target Micro-Motion Parameter Estimation}

\title{Multi-Target Micro-Motion Parameter Estimation using MIMO-FMCW Radar with Limited Measurements}

\author{Chandrashekhar~Rai*, Sanjay~J.~Alex*, and~Arpan~Chattopadhyay%
\thanks{*Co-first authors.}%
\thanks{The authors are with the Department of Electrical Engineering, Indian Institute of Technology Delhi, New Delhi, India (e-mail: eey237517@ee.iitd.ac.in; csrai.cstaff@iitd.ac.in; arpanc@ee.iitd.ac.in).}%
\thanks{The work of Chandrashekhar Rai was supported by the Visvesvaraya Post Doctoral Fellowship Scheme, Digital India Corporation, MeitY. The work of Sanjay J. Alex was supported by the AICTE Post Graduate Scholarship (GATE), All India Council for Technical Education, Ministry of Education, Government of India. The work of Arpan Chattopadhyay was supported in part by the Science and Engineering Research Board (SERB), India under Grant CRG/2022/003707, in part by the Indo-French Centre for the Promotion of Advanced Research under Grant IFC/7150/2023, and in part by Qualcomm Technologies, Inc., USA, under Project FT/2024/11/37.}}

\maketitle

\begin{abstract}
This work presents a compressive sensing-based approach for estimating the micro-motion parameters of targets with rotating components, such as small unmanned aerial vehicles (UAVs) with propellers, using fewer measurements than conventional methods. A multiple-input–multiple-output (MIMO) frequency-modulated continuous-wave (FMCW) radar employing a randomly spaced sparse linear antenna array is utilized. Random sequences of linear frequency-modulated (LFM) chirps are transmitted to enable random sampling in the slow-time domain. At first, the range, velocity, and angle of arrival (AoA) of the targets are estimated to identify the bulk motion. A three-dimensional point target response (3D-PTR) is then constructed using the estimated parameters and subtracted from the total radar return to extract the micro-Doppler signatures associated with target rotation. These residual signals are processed within a compressive sensing (CS) framework using the one-dimensional orthogonal matching pursuit (1D-OMP) algorithm to jointly estimate the propellers' rotation frequencies and blade lengths with the help of a parametric dictionary. The proposed approach is also extended to a multi-target scenario. Simulation results demonstrate that the proposed approach accurately estimates micro-motion parameters with limited measurements in the slow-time and the spatial dimensions, validating its potential for UAV detection applications. These estimated parameters are then used to distinguish among various UAV motion types using a classification framework based on a set of decision rules.

\end{abstract}
\begin{IEEEkeywords}
compressive sensing, drone detection, micro-Doppler, MIMO-FMCW radar, sparse linear arrays.
\end{IEEEkeywords}
\section{Introduction}
\label{introduction}
The use of small unmanned aerial vehicles (UAVs), particularly multirotor platforms such as quadcopters, has increased substantially over the last decade for both military and civilian applications. These platforms are now widely deployed for surveillance, logistics, infrastructure inspection, disaster management, precision agriculture, and recreational use. However, the rapid proliferation of low-cost commercial drones has also raised significant security concerns, including unauthorized surveillance, airspace violations, smuggling, and potential threats to critical infrastructure \cite{CUAS2021,Nassi2019,CISA2023}. Consequently, reliable detection, identification, and classification of UAVs have become necessary not only in military zones but also in civilian and public environments. Radar-based detection methods are at the forefront when it comes to UAV detection. However, experimental results have demonstrated that small UAVs exhibit radar cross-section (RCS) characteristics similar to those of birds \cite{Rahman2019}. Micro-Doppler signatures obtained from radar data are one of the features that help detect drones and differentiate them from birds, in addition to acoustic \cite{Shi2020, Anwar2019} and visual \cite{Zhao2022} data. The micro-Doppler effect, caused by rotating or vibrating components such as UAV propellers, introduces modulations in the target’s Doppler frequency. This effect encodes critical information about the target’s motion dynamics and structural characteristics, making it valuable for target identification and classification~\cite{Chen2003, Chen2011}. Micro-Doppler analysis also finds application in human gait movement classification \cite{Vander2018}, human vital signs detection \cite{Li2013}, ballistic target identification \cite{Gao2010}, and other related scenarios. 

\vspace{-0.4cm}
\subsection{Prior Art}
\label{Prior Art}

Time-frequency (TF) methods, such as the short-time Fourier transform (STFT), are commonly used for micro-Doppler signature analysis and parameter extraction in radar signals~\cite{Djurovic2017}. However, these methods are inherently limited by a trade-off between time and frequency resolution, which constrains their ability to precisely characterize rapidly varying micro-Doppler features. This trade-off is partially addressed by the Wigner–Ville distribution (WVD), which offers superior time-frequency resolution but suffers from cross-term interference when the signal contains more than one component~\cite{Chen1998}. To mitigate these effects, modified versions such as the pseudo-WVD (PWVD) and the smoothed pseudo-WVD (SPWVD) have been developed~\cite{Cohen1995}. These time-frequency methods have been further enhanced by integrating image processing techniques, including the Hough transform (HT) \cite{Barbarossa1996, Yang2019, Chen2000}, and the inverse Radon transform (IRT) \cite{Dakovic2013,Sharma2022}, to improve micro-motion parameter estimation. Nevertheless, these methods still face limitations, such as requiring long observation times and consequently, a large number of uniformly spaced samples.

To overcome the inherent resolution limitations of TF methods, sparse recovery techniques have been investigated for micro-motion parameter estimation in radar systems, primarily within continuous-wave (CW) settings. The sinusoidal frequency modulation–sparse recovery (SFM-SR) approach in \cite{Peng2015} demonstrated the feasibility of sparsity-based micro-Doppler estimation and is representative of early CS formulations in this domain. The pruned orthogonal matching pursuit (POMP) method in \cite{GangLi2014} focuses on estimating the micro-motion parameters of scatterers on a single target with a common rotational frequency. Similarly, \cite{Gaglione2015} employed a modified POMP algorithm to estimate helicopter propeller parameters, including the number of blades, rotation frequency, and blade length, for classification purposes. However, the analysis is restricted to a single dominant rotational frequency. The exponential orthogonal matching pursuit (EOMP) algorithm proposed in \cite{Lu2022} extended sparse modeling to estimate multiple rotational frequencies within a single UAV and incorporated machine learning (ML) techniques for flight-mode classification. Despite these advancements, the above works assume fully sampled and uniformly spaced measurements and are evaluated exclusively in single-target scenarios without addressing bulk motion.

In contrast to CW-based methods, \cite{Wang2021} employed an LFM signal model and a 3D-OMP-based sparse parametric framework to estimate the target rotational frequency, with CEMD applied for bulk body suppression. Although translational motion is addressed, the method assumes a single target with a single dominant rotational frequency. The CEMD-based suppression requires fully sampled, uniformly spaced slow-time measurements, limiting compatibility with random slow-time compressive sensing schemes. Furthermore, the approach is validated in a single-antenna setup and does not support the estimation of range, AoA, or micro-motion parameters in multi-target scenarios.
% Preserve the reference numbering used in the manually ordered manuscript:
% Peter2021, Reddy2021, Sayed2024, and Lehmann2022 must remain [26]--[29].
\nocite{Peter2021,Reddy2021,Sayed2024,Lehmann2022}
\begin{table*}[t]
\caption{Comparison of representative micro-motion parameter estimation methods.}
\centering
\label{tab:comparison}
\renewcommand{\arraystretch}{1.1}
\footnotesize
\begin{tabular}{|p{1.7cm}|p{1.7cm}|c|c|c|c|c|}
\hline
\textbf{Reference} & \textbf{Radar Type} & \textbf{Multi-Target} & \textbf{Bulk Motion} & \textbf{Range/Velocity} & \thead{\textbf{Random Slow-Time} \\ \textbf{Sampling}} & \thead{\textbf{Random Spatial} \\ \textbf{Sampling}} \\
\hline
% \cite{Peng2015}          & CW          & \xmark & \xmark & \xmark & \xmark & \xmark \\
% \hline
% \cite{GangLi2014}        & CW          & \xmark & \xmark & \xmark & \xmark & \xmark  \\
% \hline
\cite{Gaglione2015}      & CW          & \xmark & \xmark & \xmark & \xmark & \xmark\\
\hline
\cite{Lu2022}            & CW          & \xmark & \xmark & \xmark & \xmark & \xmark  \\
\hline
\cite{Wang2021}          & LFM         & \xmark & \cmark & \cmark & \xmark & \xmark  \\
\hline
\cite{Lehmann2022}       & MIMO-FMCW   & \xmark & \xmark & \cmark & \xmark & \xmark \\
\hline
\textbf{Proposed Method} & \textbf{MIMO-FMCW} & \textbf{\cmark} & \textbf{\cmark} & \textbf{\cmark} & \textbf{\cmark} & \textbf{\cmark}  \\
\hline
\end{tabular}%
\end{table*}

While most works employ pulsed or CW radar, relatively few studies have investigated the advantages of frequency-modulated continuous-wave (FMCW) radar for micro-Doppler analysis~\cite{Peter2021, Reddy2021}. It is experimentally demonstrated in \cite{Sayed2024} that multiple-input–multiple-output (MIMO) FMCW radar systems significantly outperform their single-input single-output (SISO) counterparts in UAV detection and classification, particularly under challenging angle-of-arrival (AoA) conditions. The performance gains stem from improved angular resolution and enhanced signal-to-noise ratio (SNR), which are critical for reliable spatial discrimination. Accurate AoA estimation is essential for blade-length inference due to projection effects along the radar line of sight. The use of MIMO radar for micro-Doppler analysis using TF-based methods is discussed in \cite{Lehmann2022}. Nevertheless, these approaches are developed under fully sampled, uniformly spaced acquisitions and rely primarily on TF-based processing for micro-motion characterization. Table~\ref{tab:comparison} summarizes how the proposed method compares against representative micro-motion parameter estimation approaches in the literature.

Collectively, the above limitations highlight the need for a framework that supports multi-target micro-motion analysis while jointly estimating range, velocity, and AoA, incorporating fuselage suppression, and operating under reduced measurement conditions. This motivates the development of a sparsity-driven MIMO-FMCW radar framework that enables the estimation of bulk motion parameters and micro-motion parameters using random sampling along the slow-time and spatial dimensions.

In this paper, we introduce a CS-based framework for micro-motion parameter estimation that uses a MIMO-FMCW radar system equipped with a sparse linear array (SLA) of antenna elements and transmits a non-uniform sequence of a limited number of LFM chirps within a single coherent processing interval (CPI). This architecture enables random sampling in both spatial and slow-time domains during acquisition, facilitating resource-efficient data collection. The precursor to this work is~\cite{CSrai2025}, in which the bulk motion parameter estimation algorithm is presented. In this study, those estimates are first used for target detection and bulk motion compensation, and the proposed CS-based framework is subsequently applied to multi-target micro-motion parameter estimation. The primary goal of this work is to assess whether a compressive sensing (CS)-based approach can provide more accurate UAV micro-Doppler parameter estimation than conventional time-frequency methods while requiring significantly fewer measurements. To this end, the study compares existing time-frequency analysis methods against CS-based methods within a unified framework, with the 1D-OMP algorithm selected because it is the most widely adopted and theoretically well-characterized sparse recovery algorithm, enabling a fair and interpretable comparison.

\vspace{-0.5cm}
\subsection{Our Contributions}
\label{our contribution}
To the best of our knowledge, the use of randomly transmitted sparse chirps and their integration with a random SLA for micro-motion parameter estimation of multiple targets in MIMO-FMCW radars has not been previously investigated. These aspects constitute the key novel contributions of our work, as follows:
\begin{enumerate}

\item\textbf{Modeling targets with micro-motion:}  Building upon existing micro-motion modeling approaches for small UAVs~\cite{Lehmann2022}, we develop a more comprehensive radar signal model that jointly captures both micro-Doppler modulations induced by rotating components and the bulk motion of the UAV’s fuselage. Each rotating scatterer corresponding to a propeller blade is assumed to be located at its tip, while the translational dynamics of the main body are explicitly incorporated. This integrated model provides a more realistic representation of UAV micro-motion characteristics than conventional micro-Doppler-only formulations. Existing works are limited to single-target micro-motion analysis, whereas our approach generalizes the modeling and analysis to multiple targets.
\\[0.5em] 

%\textcolor{blue}{The blade-tip scatterer model was adopted for its analytical tractability, as it yields a closed-form micro-Doppler expression that enables efficient dictionary construction for the OMP-based recovery. The primary consequence of this assumption is that the model does not account for the distributed RCS along the blade span, which would produce a smeared micro-Doppler signature rather than a discrete frequency contribution. An alternative is the thin-wire scatterer model, which integrates contributions along the blade length; however, this requires a significantly more complex and computationally expensive dictionary design. The thin-wire model is identified as a direction for future work.}
%\\[0.5em] 

\item\textbf{Extraction of micro-Doppler signatures:} The bulk motion parameters range, radial velocity, and AoA of the UAV are first estimated using the framework proposed in the precursor to this work~\cite{CSrai2025}. Based on these estimated parameters, a three-dimensional point target response (3D-PTR) is synthesized to model the radar backscatter from the UAV fuselage. This modeled contribution is then subtracted from the received radar data to suppress the bulk motion and isolate the micro-Doppler signatures arising from the rotating blades. Unlike previous works, this operation can be performed using random slow-time and spatial measurements.
\\[0.5em] 

\item\textbf{Estimation of micro-motion parameters using limited measurements:} 
The extracted micro-Doppler signatures are then used to estimate the rotational frequencies of the propellers and the effective lengths of their blades. An over-complete parameterized dictionary is constructed using the estimated Doppler frequencies and AoAs, and a sparse recovery-based parametric search is performed to obtain the corresponding micro-motion parameters. We analyze in detail how random sampling in the slow-time and spatial domains affects the performance of micro-motion parameter estimation. 
\\[0.5em] 
\item\textbf{Classification of UAV motion type:} Finally, the recovered micro-motion parameters and the estimated radial velocity are used to classify different UAV motion types or flight modes. A set of simple decision rules is applied to categorize the motion type, and the resulting classification performance is analyzed. Our framework provides an overall classification accuracy of $97\,\%$.

\end{enumerate}

The rest of the paper is organized as follows. The next section introduces the MIMO-FMCW radar system model, which accounts for targets exhibiting both bulk motion and micro-motions, and is generalized for both uniform and random measurement scenarios. Section~\ref{bulk parameter estimation} revisits the precursor work on bulk motion parameter estimation and Section~\ref{micromotionrecovery} develops the CS-based micro-motion parameter estimation method and the classification algorithm for UAV motion types. In Section \ref{simulation}, we illustrate the performance of the proposed method through extensive numerical experiments before concluding in Section~\ref{conclusion}.

Notations: Throughout this paper, bold lowercase and uppercase letters denote vectors and matrices, respectively. For a matrix $\mathbf{A}$, $[\mathbf{A}]_{:,i}$ denotes its $i$-th column, $[\mathbf{A}]_{i,:}$ its $i$-th row, and $[\mathbf{A}]_{i,j}$ its $(i,j)$-th element. The same conventions extend to higher-order arrays; for example, $[\mathbf{A}]_{i,j,h}$ denotes the $(i,j,h)$-th entry of a three-dimensional array $\mathbf{A}$.  The transpose, Hermitian, and conjugate operations are $(\cdot)^T$, $(\cdot)^H$, and $(\cdot)^*$, respectively. The Kronecker product is denoted by $\otimes$. $\lVert \cdot \rVert_{2/1/0}$ represents the $\ell_2/\ell_1/\ell_0$ norms. $\mathrm{supp}$, $\mathrm{diag}$, and $\mathrm{vec}$ denote the support, diagonal matrix formation, and vectorization operators. For a given column vector, $\mathrm{max}$ and $\mathrm{min}$ return the largest and smallest elements within that vector or set, respectively. Any quantity associated with the $k$-th target is denoted by a superscript $k$, i.e., $X^{k}$, which should be interpreted as $X$ corresponding to the $k$-th target rather than as an exponent.

\section{System Model}
\label{system model}
\subsection{Radar System Model}
\label{radar system model}

\begin{figure}[ht]
\includegraphics[width=1.05\linewidth]{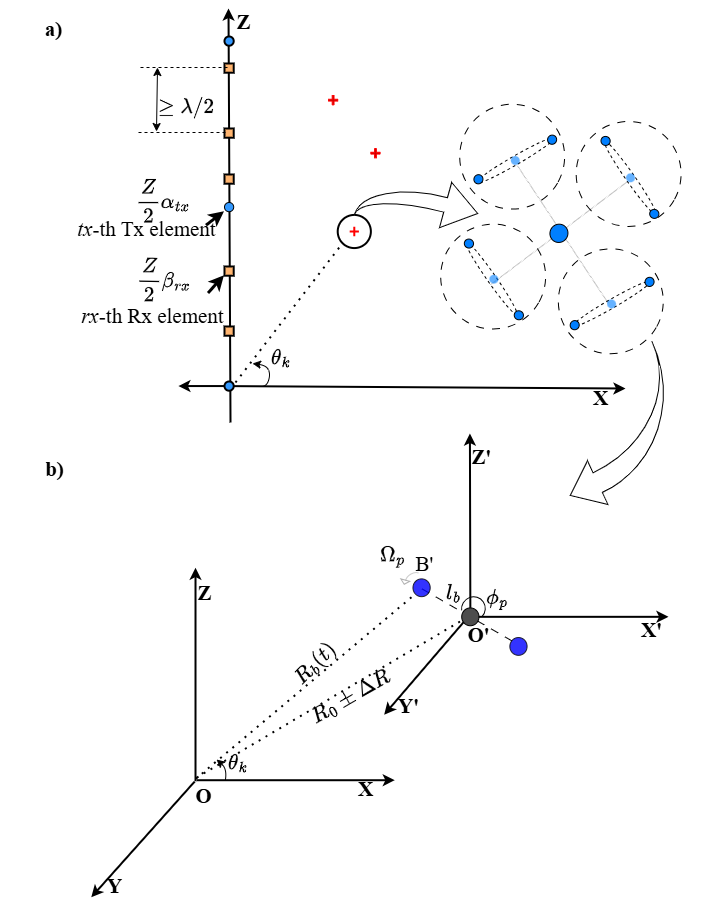}
\caption{ a) MIMO radar system ($\square$ and $\circ$ denote receivers and transmitters respectively). b) Target Geometry.}
\label{fig:fig1}
\end{figure}

In this work, we consider a co-located MIMO radar system, as shown in Fig.~\ref{fig:fig1}a, composed of \(N_{TX}\) transmitters and \(N_{RX}\) receivers forming a (possibly overlapping) array of apertures \(Z_T\) and \(Z_R\), respectively. Let us define \(Z \triangleq Z_T + Z_R\) as the total aperture. The locations of the \(\textit{tx}\)-th transmitter and \(\textit{rx}\)-th receiver along the antenna axis are \(Z\alpha_\textit{tx} / 2\) and \(Z\beta_\textit{rx} / 2\), respectively, where \(\alpha_\textit{tx} \in \left[-Z_T / Z, Z_T / Z\right]\) and \(\beta_{\textit{rx}} \in \left[-Z_R / Z, Z_R / Z\right]\). For a uniform linear array (ULA), \(\alpha_\textit{tx}\) and \(\beta_\textit{rx}\)  give uniformly distributed transmitter and receiver positions, yielding a virtual array with half-wavelength element spacing, which corresponds to the spatial Nyquist sampling rate. In the case of an SLA, \(\alpha_\textit{t}\) and \(\beta_\textit{r}\) are chosen randomly from uniform distributions, as illustrated in Fig. \ref{fig:fig1}a. We define $M$ as the total number of active virtual elements (unique transmitter--receiver position pairs formed from $\{\alpha_\textit{tx}\}$ and $\{\beta_\textit{rx}\}$), and denote their index set by $\mathcal{M} \triangleq \{m_1,m_2,\ldots,m_M\}$.

The parameters of the FMCW radar system are defined as follows. The duration of a chirp is denoted by $T_c$, the bandwidth by $B$, the chirp rate by $\gamma = B/T_c$, and the carrier frequency is represented by $f_c$. The sampling rate of the radar receiver is $f_s$, and the corresponding sampling interval is $T_s = 1/f_s$. We consider a CPI of total duration $T_p = L_{\max} T_c$, corresponding to $L_{\max}$ transmitted chirps. In the proposed sparse radar scheme, only $L_s < L_{\max}$ chirps are transmitted. Each chirp of duration $T_c$ is uniformly sampled at rate $f_s$, yielding $N = T_c/T_s$ fast-time samples per chirp. The transmitted chirp indices are denoted by $\mathcal{L}_s \triangleq \{l_1, l_2, \ldots, l_{L_s}\}$, where each $l_i$ is a distinct integer selected at random from $\{0,1,\ldots,L_{\max}-1\}$. In contrast, the traditional radar uses the full set $\{0,1,\ldots,L_{\max}-1\}$. The transmitters send orthogonal LFM chirp waveforms using time-division multiplexing (TDM), where each transmitter emits the same signal with relative time shifts. This enables the received signals from different transmitters to be easily separated at the receivers.

In our sparse setup, the time slots corresponding to the untransmitted chirps $\mathcal{L}_c = \{0, \ldots, L_{\max} - 1\} \setminus \mathcal{L}_s$, can be used to transmit chirps in other angular sectors. Throughout this work, we focus on target parameter estimation within a single angular sector. The received signals from the other sectors can be processed similarly without loss of generality. 
\vspace{-1em}
\subsection{Radar Signal Model}
\label{radar signal model}

\subsubsection{Target model}
We adopt our previously proposed radar system model in \cite{CSrai2025} and analyze the micro-motions of a target. We consider a simple target model consisting of a primary point scatterer located in the far-field, representing the rotor center, and secondary point scatterers rotating around it, representing the tip of propeller blades. Together, they govern the micro-motion characteristics of a rotor-equipped target, as shown in Fig.\ref{fig:fig1}b. To model the radar return from this configuration, we consider a 2D slant-range plane with the azimuth angle set to zero \footnote{The 2D slant-range formulation with azimuth constrained to zero is consistent with the one-dimensional sparse linear array (SLA) used in this work, which does not provide azimuth resolution. Extension to arbitrary azimuth angles would require a planar (2D) antenna array capable of simultaneous elevation and azimuth beamforming.}. We define two coordinate systems: the radar coordinate system $XOYZ$ and the target coordinate system $X'O'Y'Z'$, where $O'$ represents the center of the rotor. The rotating scatterer  $B'$, located at the tip of the blade moves around the rotor center $O'$ with an angular rotation frequency $\Omega$ radians per second on the $X'Y'$ plane. The radial distance of $B'$ from the rotor center $O'$ is $l_b$. We assume all the point scatterers denoting a target are positioned at an elevation angle $\theta$ and the distance between the radar and the rotor center is denoted by $R_0$. For small UAVs, it is safe to assume that the dimensions of the UAV are smaller than the range resolution of the radar system, so that the entire structure of the target is confined within a range bin. For example, this assumption is valid for FMCW radar systems with a carrier frequency of 24\,GHz with a 250\,MHz bandwidth, where the range resolution is approximately 60\,cm, which is larger than the full UAV structure including propeller blades like the DJI Mavic or the DJI Phantom~\cite{Lehmann2022}. Therefore, the distance from the radar to the point scatterer at the center of the propeller, as well as the bulk of the target body can be assumed to be the same, $R_0$.

Since the position of the rotating scatterer varies over time due to the blade’s rotation, the distance between $B'$ and the radar becomes time-dependent and is denoted by $R_b(t)$. The instantaneous range of the $b^{th}$ blade on the $p^{th}$ propeller of the $k^{th}$ target is given by
% \vspace{-0.25em}
\begin{equation}
\resizebox{0.98\linewidth}{!}{$
    R^k_{b,p}(t) = \sqrt{{(R^k_0)}^2 + (l^k_{b,p})^2 + 2R^k_0 l^k_{b,p}\cos(\Omega^k_{p} t + \phi^k_p)},
$}
\label{eqn:rangeonly}
\end{equation}
% \vspace{-0.25em}
where $\phi_p$ represents the initial phase offset of the blades of the $p^{th}$ propeller. Equation~\eqref{eqn:rangeonly} can be approximated by taking its binomial expansion and discarding the higher order terms of $l^k_{b,p}$/$R^k_{b,p}
$ as,

\begin{equation}
    R^k_{b,p}(t) \approx R^k_0 + l^k_{b,p} \cos(\Omega^k_{p} t + \phi^k_p).
\end{equation}

Assuming that the target's radial velocity $v^{k}$ and elevation angle $\theta^{k}$ remain invariant within a CPI, the time-dependent slant range of a rotating scatterer on the target can be expressed as

\begin{equation}
    R^k_{b,p}(t) = R^k_{0}  + v^k t +  l^k_{b,p}\cos\theta\cos(\Omega^k_{p} t + \phi^k_p).
\end{equation}

The LFM signal transmitted from the FMCW radar is

\begin{equation}
    s(t) = e^{j2\pi f_c t + j\pi \gamma t^2}   , 0<t<T_{c}.   
    \label{eqn:transmit_signal}
\end{equation}

We denote \(\tau^k_{\textit{rx},\textit{tx},l}\) as the total time delay in the radar return of the $k$-th target at the \(\textit{rx}\)-th receiver corresponding to the \(l\)-th chirp from the \(\textit{tx}\)-th transmitter. With this delay, the received signal component from $K$ such targets is given as
\vspace{-0.25em}
\begin{align}
    r_{\textit{rx},\textit{tx},l}(t) = \sum_{k =1}^{K}\tilde{a}^k \cdot s\big(t - \tau^k_{\textit{rx},\textit{tx},l}\big),\quad  
lT_c \leq t < (l+ 1) T_c ,
\label{eqn:received_signal}
\end{align}
\vspace{-0.25em}
where $\tilde{a}^k$ denotes the complex amplitude associated with the $k$-th target, proportional to its RCS. The distribution of scatterers in our signal model is such that we have a dominant scatterer on the main body of the UAV, which contributes most of the RCS, and multiple small rotating scatterers corresponding to the blades. The total time delay from the $b$-th blade on the $p$-th propeller on the $k$-th target, $\tau^k_{\textit{rx},\textit{tx},l}$, includes the delay due to radial distance and velocity $\tau_{\textit{rx},\textit{tx},l}^{k,R}$, and the delay due to the AoA $\tau_{\textit{rx},\textit{tx},l}^{k,\theta}$, which is expressed as
\begin{align}
    \tau^k_{\textit{rx},\textit{tx},l}(t) = \tau_{\textit{rx},\textit{tx},l}^{k,R}(t) + \tau_{\textit{rx},\textit{tx},l}^{k,\theta},
    \label{eqn:total_roundtripdelay}
\end{align}
\\
where
{\small
\begin{align}
    \tau_{\textit{rx},\textit{tx},l}^{k,R}(t) = \frac{2 R^k_b(t)}{c} = \frac{2 \left(R^k_0 + v^k t + l^k_{b,p} \cos\theta^k \cos(\Omega^k_p t + \phi^k)\right)}{c},
    \label{eqn:roundtrip_range}
\end{align}
}\\
and
\begin{align}
   \tau_{\textit{rx},\textit{tx},l}^{k,\theta} = \frac{Z(\alpha_\textit{tx} + \beta_\textit{rx}) \sin\theta^k}{2c}.
   \label{eqn:roundtrip_space}
\end{align}
It should be noted that the far-field assumption leads to a constant AoA across the antenna array. For the scatterer corresponding to the target fuselage, the term
\(l^k_{b,p} \cos\theta^k \cos(\Omega^k_p t + \phi^k) \) in~\eqref{eqn:roundtrip_range}
is absent, since no micro-motion is associated with the fuselage.

\subsubsection{Micro-motion analysis}
At first, for the sake of simplicity, we consider a single target and the signal contribution from a single blade of the target. After mixing the $\textit{rx}$-th received signal with the $\textit{tx}$-th transmitted waveform and applying several approximations (see~\nameref{appendixa} A), the resulting intermediate-frequency (IF) signal corresponding to one of the scatterers on the $b$-th blade can be expressed in the continuous-time domain as
\vspace{-0.25em}
{\small
\begin{align}
q_{\textit{tx},\textit{rx},l}^{b}(t) &= A^{\text{blade}} \cdot e^{j 2 \pi f_c \frac{Z (\alpha_\textit{tx} + \beta_\textit{rx}) \sin(\theta)}{2c}} \notag\label{eqn:blade_signal}
 \\
&\quad \times e^{j 2 \pi \gamma \frac{2 R_0}{c} t} 
\cdot e^{j 2 \pi \gamma \frac{2 l_b \cos\theta \cos\left(\Omega_p(t) + 2\pi \frac{b-1}{B} + \phi_p\right)}{c} t} \notag \\
&\quad \times e^{j 2 \pi f_c \frac{2 v}{c} t} 
\cdot e^{j 2 \pi f_c \frac{2 l_b \cos\theta \cos\left(\Omega_p(t) + 2\pi \frac{b-1}{B} + \phi_p\right)}{c}}. 
\end{align}
}
Here, $A^{\text{blade}}$ represents the complex amplitude of the blade return. We sample the IF signal in \eqref{eqn:blade_signal} at a rate $f_s$, predetermined by the receiver ADC, yielding \textit{fast-time} samples indexed by $n$. Sampling across chirps provides the \textit{slow-time} samples indexed by $l \in \mathcal{L}_s$, while each element of the virtual array, formed by a transmitter - receiver pair ($\alpha_\textit{tx}, \beta_\textit{rx}$), is indexed by $m \in \mathcal{M}$. Furthermore, we define the normalized Doppler and beat frequencies, along with their modulations due to the micro-Doppler effect, as follows:
{\small
\begin{equation}
    \tilde{f_v} \coloneq f_c \cdot \frac{2v T_c}{c},
\end{equation}
\begin{equation}
   \tilde{f_v}^{\mu,b,p}(l) \coloneq f_c \cdot \frac{2l_b \cos\theta \cos\left(\Omega_{p}(l T_c) + 2\pi \frac{(b-1)}{B} + \phi_p\right)}{c}, 
   \label{eqn:SFM micro frequency}
\end{equation}
\begin{equation}
\tilde{f_R} \coloneq \gamma \cdot \frac{2 R_0}{c} T_s,
\label{eqn:beat_frequency}
\end{equation}
\begin{equation}
\tilde{f_R}^{\mu,b,p}(l,n) \coloneq \gamma \frac{2l_b \cos\theta \cos\left(\Omega_{p}(l T_c) + 2\pi \frac{(b-1)}{B} + \phi_p\right)}{c}n\cdot T_s,
\end{equation}
}and the normalized spatial frequency as
\begin{equation}
  \tilde{f_a} \coloneq \frac{Z (\alpha_\textit{tx} + \beta_\textit{rx}) \sin(\theta)}{2\lambda}.  
  \label{eqn:spatial frequency}
\end{equation}

Finally, we can represent the discrete-time equivalent of~\eqref{eqn:blade_signal} as
\begin{align}
\label{eqn:Qblade}
    [\mathbf{Q}^{\text{blade}}]^{k,p,b}_{m,l,n} &=  A^{\text{blade},k} \cdot e^{j 2 \pi \tilde{f}^k_v l} 
    \cdot e^{j 2 \pi \tilde{f}_v^{\mu,k,b,p}(l)} \\
    &\quad \times e^{j 2 \pi \tilde{f}^k_R n} 
    \cdot e^{j 2 \pi \tilde{f}_R^{\mu,k,b,p}(l,n)} 
    \cdot e^{j 2 \pi \tilde{f}^k_a m}.\notag     
\end{align}
where the superscripts $k$, $p$, and $b$ denote the $k$-th target, the $p$-th propeller, and the $b$-th blade, respectively. As compared to \eqref{eqn:Qblade}, the radar return from the bulk of the target \footnote{The bulk motion is modeled such that it is not significantly affected by the micro-Doppler effect.} is modeled as
\begin{align}
    [\mathbf{Q}^{\text{bulk}}]^k_{m,l,n} = A^{\text{bulk},k} \cdot e^{j 2\pi \tilde{f}^k_v l} 
    \cdot e^{j 2\pi \tilde{f}^k_R n} \cdot e^{j 2\pi \tilde{f}^k_a m}.
    \label{eqn:Qbulk}
\end{align}

We assume that $A^{\text{blade}}$ is the same for all blades of a target and significantly lower in magnitude than $A^{\text{bulk}}$, which represents the complex amplitude of the returns from the fuselage of the target. \footnote{The uniform blade amplitude assumption ($A^{\text{blade}}$ identical for all blades) is a first-order simplification that enables analytical tractability. In practice, the RCS of each blade varies with its instantaneous orientation angle relative to the radar line-of-sight, introducing amplitude modulation across the propeller rotation cycle.} Experimental observations indicate that the relative amplitude corresponding to the RCS of the blades of small quadcopters is approximately 10–25 dB lower than that of the fuselage~\cite{Patel2018}. 

\begin{figure}[!b]
    \centering
    \subfloat[]{\includegraphics[width=0.9\linewidth]{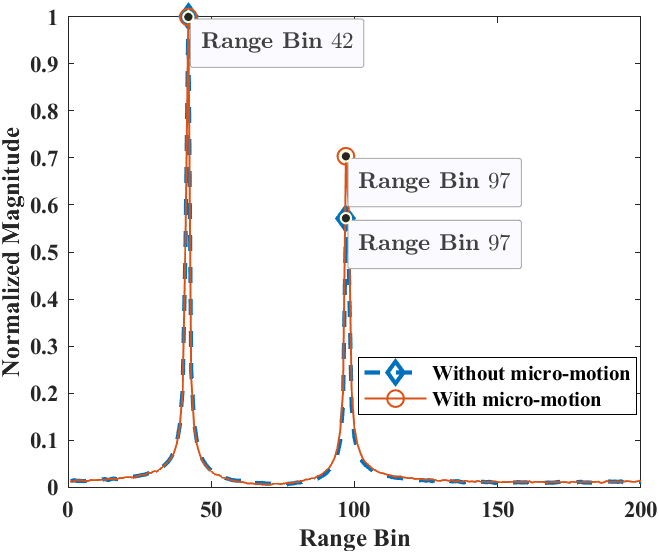}%
        \label{fig:Range_Mig}}\par
    \vspace{0.5em}
    \subfloat[]{\includegraphics[width=0.9\linewidth]{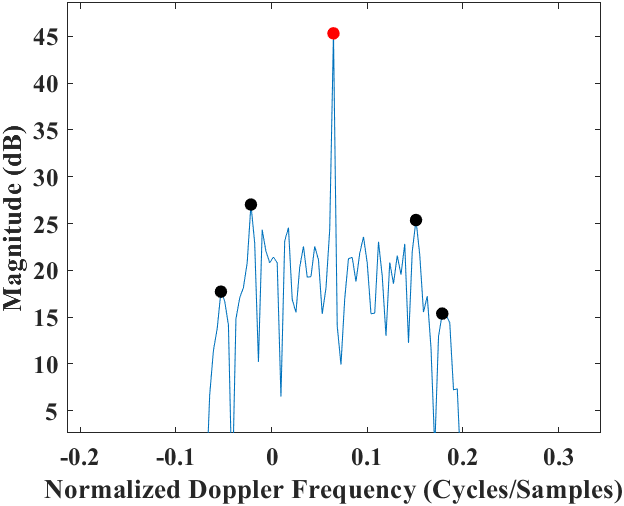}%
        \label{fig:Doppler_Mig}}

    \caption{a) Range profile of multiple targets with and without micro-motion b) Doppler characteristics in the presence of bulk and micro-motion.}
    \label{fig:Range_Doppler_Mig}
\end{figure}

Furthermore, in order to visualize the micro-Doppler effect, we analyze the phase components in the radar return of~\eqref{eqn:Qblade} under uniform sampling across all dimensions. Specifically, we compare the effects of the term $e^{j 2 \pi \tilde{f}^k_R n}$, associated with bulk target motion, and $e^{j 2 \pi \tilde{f}_R^{\mu,k}(n,l) \cdot n}$, associated with blade micro-motion, on the range profile. Our results indicate that the inclusion of the micro-motion term does not alter the range bin corresponding to the bulk motion term, as illustrated in Fig.~\ref{fig:Range_Mig}. This is attributed to the blades' limited physical dimensions ($l_b$), which are not sufficient to induce range-bin migration. Consequently, both the target body and the rotating blades remain confined within the same range bin. 

The targets under consideration consist of $P$ propellers, each comprising $B$ blades. Under the far-field assumption, the received radar signal is modeled as a superposition of backscattered signals from all blades across all propellers, along with the return from the main body. A similar decomposition framework can be found in \cite{Lu2022,Lehmann2022}.  Consequently, the received signal is modeled as the sum of the components in \eqref{eqn:Qblade} and \eqref{eqn:Qbulk} and the additive noise term, denoted by $\mathbf{W}$. Thus, the received IF signal of interest from $K$ such targets each with $P$ propellers and $B$ blades is given by,
\vspace{-0.5 cm}

\begin{align}
    \label{eqn:datacube}
    [\textbf{Q}]_{m,l,n} &= \sum_{k =1}^{K}\Big({[{\textbf{Q}^\text{bulk}]^k_{m,l,n}} +  {\sum_{p=1}^{P}}{\sum_{b=1}^B} [{\textbf{Q}^{blade}}]^{k,b,p}_{m,l,n}}\Big)\notag \\
    &\qquad+ [\mathbf{W}]_{m,l,n}.
\end{align}

 For the $k$-th target in the Doppler domain, the term $e^{j 2 \pi \tilde{f}^k_v l}$ corresponds to a single peak in the Doppler spectrum, representing the radial velocity of the complete target structure. In contrast, the terms $e^{j 2 \pi \tilde{f}_v^{\mu,k}(l)}$ introduce additional peaks and spectral spreading due to micro-motion. In the complete signal model described by \eqref{eqn:datacube}, a dominant spectral peak corresponding to the bulk Doppler frequency $\tilde{f}_v$ is observed, arising from strong radar returns associated with the fuselage and the rotating blades. This primary component (shown as the red point in Fig.~\ref{fig:Doppler_Mig}) is accompanied by symmetric side peaks (shown as black points), which result from the periodic rotational motion of the blades and characterize the micro-Doppler modulation in the received signal. These observations demonstrate that the micro-Doppler effects of small UAVs are most pronounced in the Doppler domain.

\section{Range, Velocity and AoA estimation}
\label{bulk parameter estimation}

In this section, we revisit the algorithm in~\cite{CSrai2025} for recovering the bulk target parameters i.e., range, radial velocity, and AoA from $ [\mathbf{Q}]_{m,l,n}$. The estimation of these parameters is cast as sparse recovery problems. The associated procedures are discussed in Sections~\ref{range estimation} and~\ref{jointvdoaestimation}. 

% , following the framework presented in~\cite{CSrai2025}..

\vspace{-0.5cm}
\subsection{Range Estimation}
\label{range estimation}

We perform the range estimation on the fast-time samples from a single chirp and a single receive channel in \eqref{eqn:datacube}. We can reformulate the observation model as
\begin{align}
    \mathbf{q}_r = \mathbf{D}_R \mathbf{x}_r + \mathbf{W}_r,
\end{align}
where the measurement vector $\mathbf{q}_r \in \mathbb{C}^{N \times 1}$ contains fast-time samples from a single chirp and receive channel ($\mathbf{[Q]}_{1,1,:}$). The dictionary matrix $\mathbf{D}_R \in \mathbb{C}^{N \times N_R}$, where $N_R \gg N$, is constructed using a finely discretized grid of normalized beat frequencies \(F_R=[f_{r_1}, f_{r_2}, \ldots, f_{r_{N_R}}]\), where each column corresponds to a steering vector of the form $[1, e^{j2\pi f_r}, \ldots, e^{j2\pi f_r(N-1)}]^T$. The noise vector is denoted by $\mathbf{W}_r \in \mathbb{C}^{N \times 1}$. 

We estimate the target ranges using the Range-OMP approach mentioned in \cite{CSrai2025}, which solves the following constrained optimization problem.
\begin{align}
    \min_{\mathbf{x}_r} \|\mathbf{x}_r\|_0 \quad \text{s.t.} \quad \|\mathbf{q}_r - \mathbf{D}_R \mathbf{x}_r\|_2 < \nu_1,
    \label{eqn:range_recovery}
\end{align}
where $\nu_1$ is a threshold level determined by the noise power. After recovering the support of $\mathbf{x}_r\in \mathbb{C}^{N_R \times 1}$, we obtain the estimated beat frequencies corresponding to target ranges are denoted by $\hat{f}_r^k$. The corresponding range estimates can be obtained by putting $\hat{f}_r^k$ into the equation for $\tilde{f_R}$ in \eqref{eqn:beat_frequency}.

\vspace{-1.5 em}
\subsection{Velocity and Angle Estimation}
\label{jointvdoaestimation}
Following the range estimation, we focus on estimating velocity and angle for each detected target. Let the detected range bins of the targets be $\tilde{n}_{k} = \lceil  \hat{f}^k_r\cdot N \rceil $. We select the 2D matrix over the spatial and slow-time dimensions corresponding to a particular range bin, $\tilde{n}$, as,
\begin{align}
\label{eqn:range_slice}
[\mathbf{Q}]^{\tilde{n}}_{m,l} &= 
\tilde{A}^{\text{bulk}}_{\tilde{n}} \cdot e^{j 2\pi \tilde{f}_a m} \cdot e^{j 2\pi \tilde{f}_v l} \\
&\quad + \sum_{p=1}^{P} \sum_{b=1}^{B} 
\tilde{A}^{\text{blade}}_{\tilde{n}} \cdot e^{j 2 \pi \tilde{f}_a m} \cdot e^{j 2 \pi \tilde{f}_v l} 
\cdot e^{j 2 \pi \tilde{f}_v^{\mu,b,p}(l)}\notag 
\end{align}

Here, $\tilde{A}^{\text{bulk}}_{\tilde{n}}$ and $\tilde{A}^{\text{blade}}_{\tilde{n}}$ denote the modified complex amplitudes of the target body and the blades, after the range estimation operation in Section \ref{range estimation} and the subsequent reduction to $[\mathbf{Q}]^{\tilde{n}}_{m,l}$. It also captures the contribution of spectral leakage of targets whose beat frequencies do not align with a perfect range bin.

The 2D measurement matrix corresponding to the estimated range bin $\tilde{n}$ in \eqref{eqn:range_slice} is first vectorized as $\mathbf{q}_{v,\theta} = \text{vec}([\mathbf{Q}]^{\tilde{n}}_{m,l}) \in \mathbb{C}^{ML_s \times 1}$. Let $F_a \in \mathbb{C}^{N_s \times 1}$ and $F_v \in \mathbb{C}^{N_v \times 1}$ denote the grid points of the normalized spatial and Doppler frequencies, respectively. Each point in $F_a$ is related to a potential AoA ($\theta^k$), through the relation $f_a^k = Z \sin(\theta^k)/(2\lambda)$. Similarly, each point in $F_v$ corresponds to a Doppler frequency and is related to the radial velocity $v$ by $f_v = 2vT_c/\lambda$. The dictionary used for the joint estimation is constructed as $\mathbf{D} = \mathbf{V} \otimes \mathbf{S}$, where, $\mathbf{S} \in \mathbb{C}^{M \times N_s}$, $N_s\gg M$, corresponds to the angular dictionary and is defined as
{\small
\begin{align*}
\mathbf{S} \coloneq
\begin{bmatrix}
\exp(j2\pi m_1 F_a^T) \\ \vdots \\ \exp(j2\pi m_{M }F_a^T)
\end{bmatrix},
\end{align*}}
while the matrix $\mathbf{V} \in \mathbb{C}^{L_s \times N_v}$, $N_v \gg L_s$, represents the Doppler  dictionary and is given by
{\small
\begin{align*}
\mathbf{V} \coloneq
\begin{bmatrix}
\exp(j2\pi l_{1} F_v^T) \\ \vdots \\  \exp(j2\pi l_{L_s} F_v^T)
\end{bmatrix}.
\end{align*}}

Finally, we can write the 2D measurement matrix in \eqref{eqn:range_slice} as
\begin{equation}
\mathbf{q}_{v,\theta} = \mathbf{D} \mathbf{x}_{v,\theta} + \mathbf{W}_{v,\theta} ,
\label{eqn:velocity}
\end{equation}
where $\mathbf{x}_{v,\theta}$ is a sparse vector encoding the joint Doppler–AoA components, and $\mathbf{W}_{v,\theta}$ is the noise vector. We perform the 1D-vectorized CS (OMP) method used in~\cite{CSrai2025} to recover the support of $\mathbf{x}_{v,\theta}$, which is mapped to the atoms in $\mathbf{S}$ and $\mathbf{V}$ to get the estimate of the normalized Doppler and spatial frequency of the target denoted by $\hat{f_a}$ and $\hat{f_v}$ respectively. It should be noted that recovery of parameters is dependent upon the grid size and spacing. The accuracy of the recovered parameters depends on the grid size and spacing. A comprehensive treatment of the joint Doppler- AoA estimation problem, including detailed comparisons against classical DFT, MUSIC, 1D-OMP,
2D-OMP, Basis Pursuit, and LASSO, together with an analysis of the coherence properties of $\mathbf{D} = \mathbf{V} \otimes \mathbf{S}$, is provided in~\cite{CSrai2025}. We adopt the 1D-vectorized OMP algorithm for consistency
with the preceding and subsequent processing stages, which also employ 1D-OMP in vectorized form. The corresponding velocity and AoA estimates are given by $ \hat{v} =  \frac{\lambda\hat{f_v}}{2 T_c} $, $\hat{\theta} =  \sin^{-1}\left(\hat{f}_a{} \cdot \frac{2\lambda}{Z} \right)$ respectively.

\vspace{-1em}
\section{Micro-motion parameter recovery}
\label{micromotionrecovery}
This section focuses on the recovery of micro-motion parameters. Accurate estimation of these parameters requires isolating the micro-motion signatures from the dominant bulk response. To this end, we first cancel out the contribution of the bulk motion using the estimated range, radial velocity, and AoA. This enables the extraction of the residual signal containing primarily micro-Doppler components, as described in Section~\ref{extractionmd}. The extracted micro-Doppler signatures are then used in a sparse parametric framework to jointly estimate the micro-motion parameters, including blade length and rotational frequency, as detailed in Section~\ref{micromotion estimation}.
\subsection{Extraction of Micro-Doppler Signatures}
\label{extractionmd}

The next step is to extract the micro-motion components from the radar return in \eqref{eqn:datacube}. This requires removing the contribution of the bulk motion, whose relatively large complex amplitude ($\tilde{A}_{\text{bulk}}$) can obscure micro-Doppler signatures and lead to inaccurate micro-motion parameter estimation. 

To accurately estimate $\tilde{A}_{\text{bulk}}$, we first construct the basis vectors  $\tilde{\mathbf{v}}(\hat{f_v}) \coloneq \left[\exp(-j2\pi l_1\hat{f_v}), \ldots, \exp(-j2\pi l_{L_s}\hat{f_v})\right]^{\mathrm{T}}, \text{and} \quad \tilde{\mathbf{s}}(\hat{f_a}) \coloneq \left[\exp(-j2\pi m_1\hat{f_a}), \ldots, \exp(-j2\pi m_M\hat{f_a})\right]^{\mathrm{T}}$.
We then perform a correlation operation on  $\left[\mathbf{Q}\right]^{\tilde{n}}_{m,l}$ in \eqref{eqn:range_slice} with $\tilde{\mathbf{v}}$ along the slow-time dimension and $\tilde{\mathbf{s}}$ along the virtual array dimension. This operation can be viewed as a generalization of the familiar one-dimensional inner product to two dimensions. Extending this idea, $\left[\mathbf{Q}\right]^{\tilde{n}}_{m,l}$ is projected onto a pair of basis vectors, one for each axis (i.e., spatial and slow-time). This two-dimensional projection calculates the contribution of the rank-1 structure formed by the outer product of the two bases within the matrix. Mathematically, this is expressed as

\vspace{-0.1em}
\begin{equation}
\label{eqn:2D_Projection}
\hat{\tilde{A}}^{\text{bulk}} = \frac{\tilde{\mathbf{s}}^{T} \left[\mathbf{Q}\right]^{\tilde{n}}_{m,l} \tilde{\mathbf{v}}}{N \cdot M \cdot L_s}.
\end{equation}

Further, we construct a 3D-PTR using $\hat{\tilde{A}}^{\text{bulk}}$, $\hat{f_v}$, $\hat{f_r}$, and $\hat{f_s}$, to model the contribution of the bulk motion in the radar IF signal in \eqref{eqn:datacube}. Therefore, we define the estimated 3D-PTR as
\vspace{-0.2em}
\begin{align}
\label{eqn:PSF}
\mathbf{[Q]}^{\text{PTR}}_{m,l,n} &\coloneq \hat{\tilde{A}}^{\text{bulk}} \cdot e^{j2\pi \hat{f_v} l} \cdot e^{j2\pi \hat{f_r} n}\cdot e^{j2\pi \hat{f_s} m}.
\end{align}

We then extract the micro-motion components by subtracting the estimated bulk response, defined by the PSF in~\eqref{eqn:PSF}, from the original radar IF signal in~\eqref{eqn:datacube}. The resulting residual signal, denoted $\tilde{\mathbf{[Q]}}^{\mu}_{m,l,n}$, contains primarily the micro-Doppler contributions.
\vspace{-0.2em}
\begin{align}
\label{eqn:CLEAN}
\tilde{\mathbf{[Q]}}^{\mu}_{m,l,n} = \mathbf{[Q]}_{m,l,n} - \mathbf{[Q]}^{\text{PTR}}_{m,l,n}.
\end{align}

\begin{figure}[!htbp]
\vspace{-0.2cm}
    \centering
    \subfloat[Doppler spectrum with bulk component.]{%
        \includegraphics[width=0.47\linewidth]{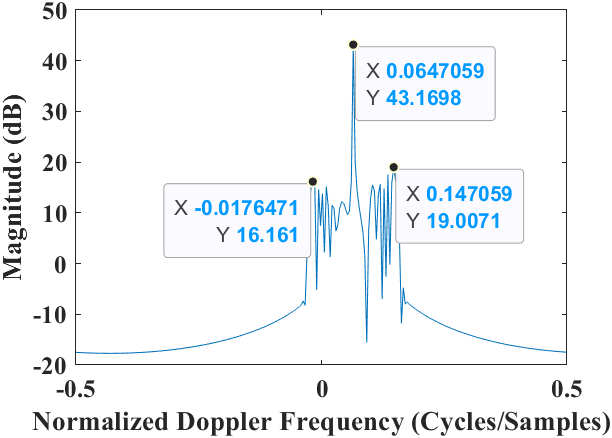}}
    \hfill
    \subfloat[Doppler spectrum after removing the bulk component.]{%
        \includegraphics[width=0.47\linewidth]{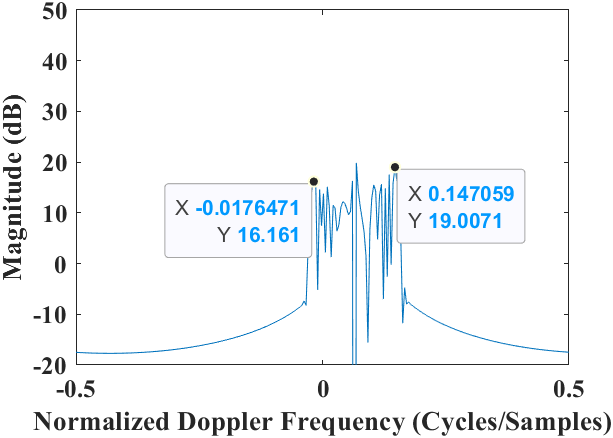}}

    \caption{Extraction of Micro-Doppler Signatures.}
    \label{fig:PSF_Sub}
\end{figure}

This process is illustrated in Fig.~\ref{fig:PSF_Sub},where the magnitude of the complex amplitude is plotted against the normalized Doppler frequency. For illustrative purposes, uniform sampling across the full set of chirps within a CPI is employed. However, the proposed method applies equally well when using a reduced set of randomly selected samples. In Fig.~\ref{fig:PSF_Sub}(a), the Doppler spectrum exhibits a dominant peak attributed to the target’s bulk motion (non-zero radial velocity). Fig.~\ref{fig:PSF_Sub}(b) shows that this peak is effectively suppressed, revealing side peaks corresponding to micro-Doppler modulations. 

%\textcolor{blue}{The bulk parameters for the 3D-PTR, i.e., the range, Doppler, and AoA parameters, are estimated with inherent estimation error as characterized in the precursor work~\cite{CSrai2025}, and are therefore not assumed to be perfectly known. Consequently, the 3D-PTR does not completely cancel the bulk-motion component; rather, it suppresses it so that the micro-Doppler features become prominent, thereby aiding the recovery of the micro-motion parameters. To reflect these non-ideal conditions, the initialization of the bulk parameters in the simulations is random rather than deterministic, ensuring that the framework is evaluated in the presence of bulk-parameter estimation errors. The simulations in Section\ref{simulation} account for estimation error through random initialization. }

\subsection{Micro-motion Parameter Estimation}
\label{micromotion estimation}

Further, to analyze the micro-Doppler signatures, $\tilde{\mathbf{Q}}^{\mu}$ is sliced along the range dimension at a specific range bin $\tilde{n}$, obtained after the range estimation in Section~\ref{range estimation}. The result is a 2D matrix $[\tilde{\mathbf{Q}}^{\mu}]_{m,l} \doteq [\tilde{\mathbf{Q}}^{\mu}]_{m,l,n}\big|_{\tilde{n}}$ which spans the spatial ($m$) and slow-time ($l$) dimensions. To compensate for the spatial frequency, the matrix $[\tilde{\mathbf{Q}}^{\mu}]_{m,l}$ is multiplied by the steering vector $\tilde{\mathbf{s}}$, defined in Section~\ref{extractionmd}. This operation effectively cancels the spatial component in $\tilde{\mathbf{Q}}^{\mu}_{m,l}$ resulting in a vector only in the slow-time domain as
\begin{equation}
\label{eqn:spatial_proj}
    \tilde{\mathbf{q}}^{\mu}_l = (\tilde{\mathbf{Q}}^{\mu }_{m,l})^{T} \tilde{\mathbf{s}} \quad  \in \mathbb{C}^{ L_s \times 1} .
\end{equation}

The column vector $\tilde{\mathbf{q}}^{\mu}_l \in \mathbb{C}^{ L_s \times 1}$, contains the micro-Doppler variations over slow-time, after removing the spatial and the dominant Doppler component associated with bulk motion. The underlying micro-motions in $\tilde{\mathbf{q}}^{\mu }_l$ can be modeled as an SFM signal \cite{Peng2014}, which effectively models the periodic Doppler fluctuations caused by rotating or vibrating structures such as propeller blades. The SFM signal space is defined as
\begin{align}
\mathcal{S} \coloneq \Bigg\{ s(t) \;\Bigg|\; 
s(t) = \sum_{i} \exp\!\left( j a_{i} \cos \!\left(\omega_{i} t + \varphi_{i}\right) \right),\\\; \notag
a_{i} \in \mathbb{C},\;
\omega_{i} \in \mathbb{R},\;
\varphi_{i} \in [0,2\pi)
\Bigg\}.
\end{align}
Here, $a_{i}$, $\omega_{i}$, and $\varphi_{i}$ represent the amplitude, angular frequency components, and initial phase angle, respectively. The vector $\tilde{\mathbf{q}}^{\mu }_l$ is essentially a combination of  frequency components of the form $\tilde{f}_v^{\mu,k}(l)$ defined in \eqref{eqn:SFM micro frequency}. Therefore, it can be modeled as a member of the signal family $\mathcal{S}$, characterized by a sum of complex exponentials with sinusoidal phase modulation. In this representation, only a few parameters are required to accurately describe $\tilde{\mathbf{q}}^{\mu }_l$, implying its sparsity in the SFM domain. The proposed framework exploits this sparsity characteristic to enable efficient parameter estimation.

We construct a parameterized dictionary \(\mathbf{\tilde{F}}_{\mathrm{mD}}\) using $\hat{f}_v$ and $\hat{\theta}$ for the joint estimation of the blade length \(l_b\), rotation frequency \(\Omega\), and initial phase \(\phi\) from the support of a sparse vector. We define the discretized parameter grids for \(l_b\), \(\Omega\), and \(\phi\) as \(\mathcal{I}\), \(\mathcal{W}\), and \(\mathcal{E}\), respectively. Each atom \(\psi_{r,o,\epsilon}\) of \(\mathbf{\tilde{F}}_{\mathrm{mD}}\) corresponds to a unique triplet of blade length, rotational frequency, and initial phase indexed by \(r\text{,}\,\,o \, \, \text{and} \, \,\epsilon\), respectively and is defined as
\begingroup
\large
\vspace{-0.25em}
\begin{align}
    \psi_{r,o,\epsilon}(l) &\coloneq e^{{j 2\pi \hat{f}_v^k l}}\\
    &\hspace{-4em} \times \sum_{b=1}^Be^{\Big( j \frac{4\pi}{\lambda} \mathcal{I}(r) \cos(\hat{\theta}^k) 
    \cos(2\pi \mathcal{W}(o) l + \mathcal{E}(\epsilon)+ 2\pi \frac{(b-1)}{B} )\Big)}.\notag
    % \label{eq:atom_phi}
\end{align}
\endgroup
 % Prior knowledge of the possible range of values of blade length, rotation frequency and number of blades is required to construct $\mathbf{\tilde{F}}_{\mathrm{mD}}$.
 
 Let \(\Gamma\), \(\Lambda\), and \(\Upsilon\) denote the total number of grid points in \(\mathcal{I}\), \(\mathcal{W}\), and \(\mathcal{E}\), respectively. The over-complete  dictionary is formed by concatenating all atoms across the parameter grid as,
\vspace{-0.5em}
\begin{align}
    \mathbf{\tilde{F}}_{\mathrm{mD}} &= [\mathbf{F}_{\mathrm{mD}}(0,0,0), \dots, \mathbf{F}_{\mathrm{mD}}(\Gamma-1,0,0),\dots,\\
    \hspace{-2em}&\hspace{-1.5em}\mathbf{F}_{\mathrm{mD}}(0,\Lambda-1,\Upsilon-1), \dots
    \mathbf{F}_{\mathrm{mD}}(\Gamma-1,\Lambda-1,\Upsilon-1)]\notag,
\end{align}
\vspace{-0.5em}
where each atom \(\mathbf{F}_{\mathrm{mD}}(r,o,\epsilon)\) is defined as,
\begin{align}
    \mathbf{F}_{\mathrm{mD}}(r,o,\epsilon) &= [\psi_{r,o,\epsilon}(0), \dots, \psi_{r,o,\epsilon}(L_s - 1)]^T.
\end{align}
 Thus, the dictionary \(\mathbf{\tilde{F}}_{\mathrm{mD}} \in \mathbb{C}^{L_s \times (\Gamma  \Lambda \Upsilon)}\) contains all candidate atoms required for joint estimation of \(l_b\), \(\Omega\), and \(\phi\). Finally,  $\tilde{\mathbf{q}}^{\mu}_l$ can be reformulated as, 
\begin{align}
\label{eqn:microDoppler}
\tilde{\mathbf{q}}_{l}^{\mu} &= \mathbf{\tilde{F}}_{\mathrm{mD}} \mathbf{x}_{\mathrm{mD}} + \mathbf{W}_{\mathrm{mD}}, 
\end{align}
where $\mathbf{W}_{\mathrm{mD}}$ represents the noise vector. Specifically, the observed vector $\tilde{\mathbf{q}}_{l}^{\mu}$ is modeled as a linear combination of candidate atoms, and the unknown coefficient vector $\mathbf{x}_{\mathrm{mD}} \in \mathbb{C}^{\Gamma \cdot \Lambda \cdot \Upsilon \times 1} $ is sparse. Following the approach in Section~\ref{jointvdoaestimation}, the parameter estimation task is reduced to a one-dimensional sparse recovery problem as, 
\begin{align}
\label{eqn:microdoppler_recover}
    \min_{\mathbf{x}_{\mathrm{mD}}} \quad & \|\tilde{\mathbf{q}}^{\mu}_l - \mathbf{\tilde{F}}_{\mathrm{mD}} \mathbf{x}_{\mathrm{mD}}\|_2, \;  \text{s.t} \quad \| \mathbf{x}_{\mathrm{mD}}\|_0 =  P.
\end{align}

\begin{algorithm*}[!htbp]
\caption{Compressive Sensing-Based Micro-Motion Parameter Estimation}
\label{alg:algorithm1}
\begin{algorithmic}[1]
\State \textbf{Input:} Radar IF data $[\textbf{Q}]_{m,l,n}$.
\State Perform range estimation on $[\textbf{Q}]_{m,l,n}$ using the Range-OMP algorithm in Section~\ref{range estimation} to obtain $\tilde{f}_r^k$.
\For{each detected range bin $k = 1, 2, \dots, \tilde{K}$}
    \State \parbox[t]{425pt}{%
        Extract the 2D spatial-slow-time matrix corresponding to the detected range bin as shown in \eqref{eqn:range_slice}.}
    \State \parbox[t]{425pt}{Perform joint estimation of Doppler $(\tilde{f}_v^k)$ and spatial frequencies $(\tilde{f}_a^k)$ as mentioned in Section~\ref{jointvdoaestimation}.%
    }
    \State \parbox[t]{425pt}{%
        Estimate  $\hat{A}_{\text{bulk}}^k$, by performing the 2D projection operation on $[\mathbf{Q}]^{\tilde{n}}_{m,l}$ as shown in \eqref{eqn:2D_Projection}.%
    }
    \State \parbox[t]{425pt}{%
        Construct the 3D-PTR using $\tilde{f}_r^k , \tilde{f}_v^k, \tilde{f}_a^k \; \& \;\hat{A}_{\text{bulk}}^k $ via \eqref{eqn:PSF}.
    }
    \State \parbox[t]{425pt}{%
        Subtract the 3D-PTR from $\textbf{[Q]}_{m,l,n}$ to suppress the contribution of the bulk as in \eqref{eqn:CLEAN}. 
    }
    \State \parbox[t]{425pt}{%
        Obtain the slow-time signal $\tilde{\mathbf{q}}^{\mu }_l$ as shown in  \eqref{eqn:spatial_proj}.
    }
    \State \parbox[t]{425pt}{%
        Construct a parametrized dictionary  $\mathbf{F_{mD}}$ using $ \tilde{f}_v^k\; \& \;\tilde{\theta}^k $. 
    }
    \State \parbox[t]{425pt}{%
         Solve \eqref{eqn:microdoppler_recover} using 1D-OMP Algorithm to obtain P prominent peaks corresponding to the number of detected propellers.}
         
    \State \parbox[t]{425pt}{Each peak corresponds to  a unique triplet of micro-motion parameters $\{\hat{\Omega}^k, \hat{l}_b^k, \hat{\phi}^k\}$.
    }
\State \parbox[t]{400pt}{\textbf{Output:} Micro-motion parameters $\hat{\mathbf{\Omega}}$ (as a set of rotation frequencies) and $ \hat{l}_b$ along with range, velocity and AoA estimates for each target.}
\EndFor

\end{algorithmic}
\end{algorithm*}

 We apply the 1D-OMP algorithm to recover the support of $\mathbf{x}_{\mathrm{mD}}$ which gives us $P$ non-zero elements. Index of each nonzero element in $\mathbf{x}_{\mathrm{mD}}$ maps uniquely to a point on the discretized parameter grid space, $\mathcal{I} \times \mathcal{W} \times \mathcal{E}$. These grid indices corresponding to the identified support are then mapped directly to the corresponding physical parameters: \( l_b\),  \(\Omega\), and  \( \phi\), based on the grids of $\mathcal{I}$, $\mathcal{W}$, and $\mathcal{E}$ respectively. As a note, although the initial phase angle is estimated by the algorithm, it serves as a nuisance parameter rather than a physically meaningful target attribute. However, its inclusion in the model is necessary as it influences the recovery of \( \Omega \) and \( l_b \). The steps from Section~\ref{jointvdoaestimation} are repeated for all the estimated range bins in case of multiple detected targets. The overall algorithm is summarized in Algorithm~\ref{alg:algorithm1}.

\subsection{Computational complexity}
\label{computational complexity}
The range estimation has a computational complexity of the order of $\mathcal{O}(KN_RN)$. The joint estimation of velocity and AoA has a computational complexity on the order of $\mathcal{O}(N_sN_vML_s)$. The construction of the 3D-PTR and its subsequent removal have a computational complexity of the order $\mathcal{O}(NML_s)$. The micro-motion parameter estimation of a single target has a computational complexity of $\mathcal{O}(PL_s\Gamma  \Lambda \Upsilon )$. Thus, the overall computational complexity is $\mathcal{O}\left(KN_RN + N_sN_vML_s + NML_s + PL_s \Gamma \Lambda \Upsilon \right)$.

\subsection{Classification of UAV motion type}

The proposed framework enables not only propeller rotational frequency estimation but also quadcopter flight mode classification. This capability is achieved through a rule-based classifier that makes use of the estimated propeller frequencies ({$\hat{\Omega}$}) and the radial velocity ($\hat{v}$) of the target. The key quantities used in the classification process are defined as follows. The \textit{estimated rotor frequency vector} $\hat{\Omega}$ comprises the estimates of rotational frequencies of all $P$ propellers. The \textit{mean rotational frequency} denoted by $\bar{\Omega} = \frac{1}{P} \sum_{i=1}^{P} \hat{\Omega}_i$ represents the average rotation frequency and serves as a reference for detecting collective variations in rotational dynamics. The \textit{front--rear frequency difference} $\bar{\Omega}_{fr}$ measures the difference between the frequencies of the front and rear propellers, providing an indicator of asymmetric thrust that occurs during forward or backward translational motion. The \textit{hover reference frequency} $\Omega_h$ corresponds to the nominal propeller rotational speed during stationary hover. The classification algorithm is summarized in Algorithm~\ref{alg:classifier}.
\begin{algorithm}[]
\caption{Quadcopter Flight Mode Classification}
\label{alg:classifier}
\begin{algorithmic}[1]
\State \textbf{Input:} Estimated rotor frequency vector $\hat{\Omega}$ sorted in ascending order, mean rotational frequency $\bar{\Omega}$, 
front--rear frequency difference $\bar{\Omega}_{fr}$, estimated velocity $\hat{v}$, 
and hover reference frequency $\Omega_h$.
\vspace{4pt}
\State \textbf{Threshold parameters:} Maximum frequency spread threshold $\Delta_{\Omega}$, 
hover deviation threshold $\Delta_{h}$, higher 
front--rear differential threshold  $\Delta_{fr,h}$, lower 
front--rear differential threshold  $\Delta_{fr,l}$, 
and velocity limits $v_{\text{low}}$, $v_{\text{high}}$.
\vspace{4pt}
\If{$\max(\hat{\Omega}) - \min(\hat{\Omega}) < \Delta_{\Omega}$ \textbf{and} $|\bar{\Omega} - \Omega_h| < \Delta_{h}$ \textbf{and} $|\hat{v}| < v_{\text{low}}$}
    \State Flight mode $\gets$ \textbf{Hover}
\ElsIf{($|\bar{\Omega}_{fr}| > \Delta_{fr,h}$ \textbf{and} $|\hat{v}| > v_{\text{low}}$) \textbf{or} $|\hat{v}| > v_{\text{high}}$}
    \State Flight mode $\gets$ \textbf{Translation}
\ElsIf{$|\bar{\Omega} - \Omega_h| > \Delta_{h}$ \textbf{and} $|\bar{\Omega}_{fr}| < \Delta_{fr,l}$}
    \If{$\bar{\Omega} > \Omega_h$}
        \State Flight mode $\gets$ \textbf{Takeoff}
    \Else
        \State Flight mode $\gets$ \textbf{Landing}
    \EndIf
\Else
    \State Flight mode $\gets$ \textbf{Hover}
\EndIf
\State \textbf{Output:} Flight mode
\end{algorithmic}
\end{algorithm}

\section{Simulation Results}
\label{simulation}
We conduct numerical experiments using synthetic data generated as per the radar system model in Section~\ref{radar system model}  and the radar signal model of Section~\ref{radar signal model}. The radar parameters used for the simulations are presented in Table~\ref{tab:parameters}. The results of bulk motion parameter estimation in this work are consistent with those reported in~\cite{CSrai2025}, and are therefore not repeated here. All simulations are conducted on a workstation running MATLAB R2024a with an Intel Core i7 processor (3.6 GHz, 8 cores) and 16 GB of RAM.

\begin{table}[H]
\centering
\setlength{\tabcolsep}{5pt} % adjust column spacing
\renewcommand{\arraystretch}{0.9} % adjust row spacing
\footnotesize
\begin{tabular}{|l|c|c|}
\hline
\textbf{Description} & \textbf{Parameter} & \textbf{Value} \\ \hline
Sampling frequency            & $f_s$       & $5 $ MHz \\
Carrier frequency             & $f_c$       & $24 $ GHz \\
Carrier wavelength             & $\lambda$       & $12.5 $ mm \\
Bandwidth of operation        & $B$         & $250$ MHz \\
Chirp duration                & $T_c$       & $40$ $\mu\text{s}$ \\
Chirp rate                    & $\gamma$    & $6.25 \times 10^{12}$\\
Number of chirps              & $L$         & $64 - 512$ \\
Aperture length               & $Z$         & $6\lambda$ \\
No. of antenna elements (SLA) & $N_T, N_R$  & $2, 4$ \\
\hline
\end{tabular}
\caption{Radar system parameters.}
\label{tab:parameters}
\vspace{-0.3cm}
\end{table}

\subsection{Experimental Design}
The performance of the proposed algorithm is evaluated using 500 Monte Carlo simulations for each combination of SNR and compression ratio. The specified SNR values are applied to the extracted micro-Doppler signatures after removing the bulk motion component associated with the target’s translational movement. The compression ratio (CR) is defined as the ratio of the number of randomly selected chirps $L_s$ used in a CPI to the total number of chirps available for the full CPI $L_{\max}$, i.e., $\text{CR} = L_s / L_{\max}$. We have also included uniform sampling in the slow time axis (denoted as "uniform" in the plots) for comparison. Uniform sampling in slow-time refers to using the full set of $L_{max}$ chirps in a CPI for a chosen angular sector, with chirp transmissions occurring at uniformly spaced time intervals. The term random sampling is used when only a subset of chirps is selected according to a particular CR, resulting in a reduced subset of non-uniformly spaced slow-time samples. Throughout the experiments we use a $2 \text{Tx} \times 4 \text{Rx} $ SLA in which the positions of the array elements are randomly selected in each trial resulting in random sampling in the spatial domain. In each trial bulk target parameters including range, bulk velocity, AoA, complex amplitude of the fuselage and micro-motion parameters (propeller rotational frequencies \(\Omega\), blade length \( l_b \), and phase angle \(\phi\)) are initialized independently from uniform distributions whose bounds are consistent with the assumed signal model.

\begin{figure}
    \centering
    \includegraphics[width=0.75\linewidth, height=5cm]{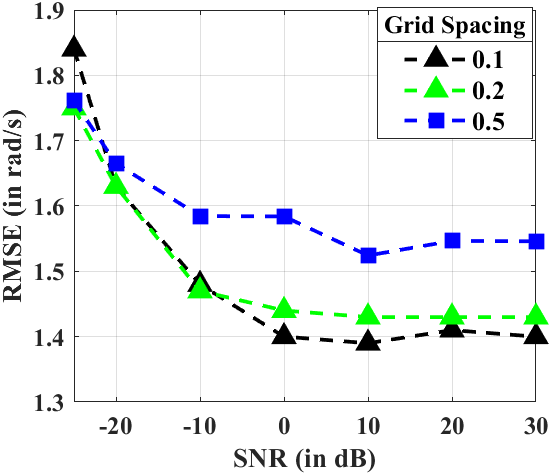}
    \caption{RMSE of rotational angular frequency $(\Omega)$  vs SNR for different grid spacings for $\mathcal{W}$.}
    \label{fig:grid_resolution}
\end{figure}

The micro-Doppler dictionary \( \mathbf{\tilde{F}}_{\mathrm{mD}} \), is generated using discretized parameter sets $\mathcal{I} \times \mathcal{W} \times \mathcal{E}$. Specifically, the grid of blade lengths $\mathcal{I}$ is uniformly sampled over $[0.1, 0.2]~\text{m}$ with 20 points and the grid of phase offset values $\mathcal{E}$ is uniformly discretized into 20 points spanning $[0, \pi]~\text{rad}$ as one of the two scatterers on a blade will have an initial phase in this range and the phase of the other is at $\pi~\text{rad}$ from the other.
The grid of rotational frequency values $\mathcal{W}$ is defined over the range \([50, 90]~\text{rps}\), consistent with the motion characteristics of small commercial UAVs~\cite{Deter2017}. A grid spacing of $0.1~\text{rps}$ is selected for $\mathcal{W}$ to achieve a suitable trade-off between estimation accuracy and computational cost based on a root mean squared error (RMSE) vs SNR analysis across multiple grid spacings for $\mathcal{W}$ with the other micro-motion parameters kept constant as shown in Fig.~\ref{fig:grid_resolution}. Additionally, the grids for range frequency \( F_R \), Doppler frequency \( F_v \), and spatial frequency \( F_a \) are discretized uniformly within the intervals \([0, 1]\), \([-0.5, 0.5]\), and \([-0.25, 0.25]\) with a granularity of 0.005 for all three. For \( F_a \), these grids correspond to an angular sector of approximately \(-40^\circ\) to \(40^\circ\) in elevation.

Off-grid effects arise when the true parameter values do not coincide with the grid points of the CS dictionary, causing energy to leak across adjacent atoms and degrading estimation accuracy. In this work, the dictionaries are constructed on a fixed grid that is independent of the target parameters, while the true parameters (range, velocity, AoA, and micro-motion frequencies) are drawn from continuous distributions. Consequently, the true values almost surely do not fall on grid points, and every Monte Carlo trial reflects an off-grid condition. The impact of this mismatch is quantified explicitly in the RMSE-versus-grid-resolution analysis, which characterizes how the residual off-grid leakage contributes to the overall error as the grid is refined as seen in Fig.\ref{fig:grid_resolution}. The off-grid problem for OMP-based micro-motion parameter recovery is well studied in \cite{GangLi2014}.

\vspace{-0.5em}
\subsection{Performance Metrics}
\label{performance metrics}
\vspace{-0.5em}

The performance of the proposed detection framework is evaluated using the hit rate and RMSE. Consider an experimental setup comprising $K$ targets, each equipped with $P$ propellers and $B$ blades. In each trial, the total number of true propeller rotation frequencies is $K \times P$.  The hit rate is defined as the ratio of the number of correctly estimated rotation frequencies (those within a tolerance of $1.25$~rps of their corresponding true values) to the total number of true frequencies.  The accuracy of the estimated parameters is quantified using the RMSE as defined in~\eqref{eq:rmse}. Here, $\hat{e}_i$ and $e_i$ denote the estimated and true scalar values of a micro-motion parameter, respectively and $N_\text{hits}$ represents the total number of successful detections (hits) across all trials.
\begin{equation}
\mathrm{RMSE} = \sqrt{\frac{1}{N_\text{hits}} \sum_{i=1}^{N_\text{hits}} (\hat{e}_i - e_i)^2}
\label{eq:rmse}
\end{equation}

\begin{figure}[!t]
    \centering
    \includegraphics[width=0.75\linewidth, height=5cm]{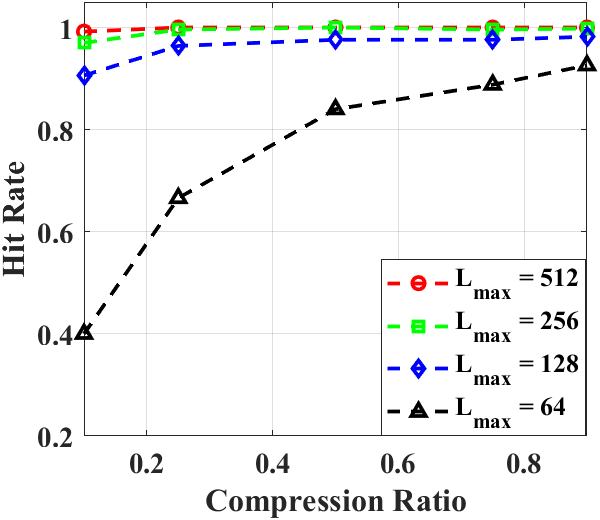}
    \caption{Hit Rate vs Compression Ratio  for SNR = 20 dB.}
    \label{fig:HRvsCR}
\end{figure}

The noise term is modeled as zero-mean complex additive white Gaussian noise. For each trial, the signal power is computed empirically, and the noise variance is derived by scaling this signal power according to the target SNR (specified in dB), ensuring the desired SNR is achieved.Since both metrics are estimated from a finite number of Monte Carlo trials $N$ at each SNR point, 95\% confidence intervals are also reported to indicate their statistical reliability. For the hit rate, the interval is computed using the Wilson score approximation for a binomial proportion, which remains well-behaved even as the hit rate approaches 0 or 1. For the RMSE, assuming approximately Gaussian estimation errors, the interval follows from the fact that $N_\text{hits}(\mathrm{RMSE})^2/\sigma^2$ is $\chi^2$-distributed with $N_\text{hits}$ degrees of freedom, where $\sigma$ is the true underlying error magnitude being estimated. These intervals are shown as error bars in the corresponding results plots.

\subsection{Finding the Suitable $L_{max}$}
\label{suitable lmax}

At first, simulations are performed for different numbers of chirps (\( L_{max} = 64, 128, 256, 512\)) and a $2 \text{Tx} \times 4 \text{Rx} $ SLA. For each Monte Carlo run, the array elements are placed randomly in an aperture of physical length $6\lambda$ with their positions chosen from a uniform distribution. For each value of \(L_{max}\), different \(L_s\) values are used corresponding to CRs ranging from \(0.1 \) to \(0.9\) with the SNR fixed at \(30\,\mathrm{dB}\). The results for a single target with a single propeller indicate that \( L_{max} = 512 \) and \( L_{max} = 256 \) yield comparable performance, with both outperforming smaller values of \(L_{max} \) as observed in Fig.\ref{fig:HRvsCR}. Hence,  \( L_{max}= 256 \) is chosen for the remaining experiments to ensure reduced sampling cost without compromising performance.

\vspace{-0.5em}

\subsection{Performance Analysis}

Further evaluations of the proposed method are conducted for both single-target (\(K = 1\)) and dual-target (\(K = 2\)) configurations. The antenna array configuration is identical to that described in Section~\ref{suitable lmax}. Each target is modeled with \(P = 2\) and \(P = 4\) propellers, respectively, while all experiments are performed with a fixed number of blades \(B = 2\). Although the proposed algorithm can operate when the number of propellers is unknown by imposing a nominal sparsity level of $P = 4$ in~\eqref{eqn:microdoppler_recover} for both $P = 2$ and $P = 4$ target scenarios, this overestimation of the true sparsity level leads to reduced estimation accuracy. Specifically, for a target with two propellers, enforcing a sparsity constraint of $P = 4$ results in recovery of the two dominant peaks corresponding to the true propellers along with additional spurious peaks of significantly lower magnitude in their vicinity. These spurious components can be removed through appropriate magnitude thresholding. However, the overall estimation accuracy is still degraded.  Therefore, we assume that both the number of propellers and blades are known for the performance analysis. We  evaluate the performance based on the hit rate and the RMSE of the estimated angular rotation frequency vector (\(\hat{\mathbf{\Omega}}\)) and blade length (\(l_b\)) under varying SNR conditions for two different cases as follows.

\begin{figure*}[!t]
    \renewcommand{\thefigure}{7}
    \centering
    \subfloat[]{\includegraphics[width=0.32\textwidth,height=4cm]{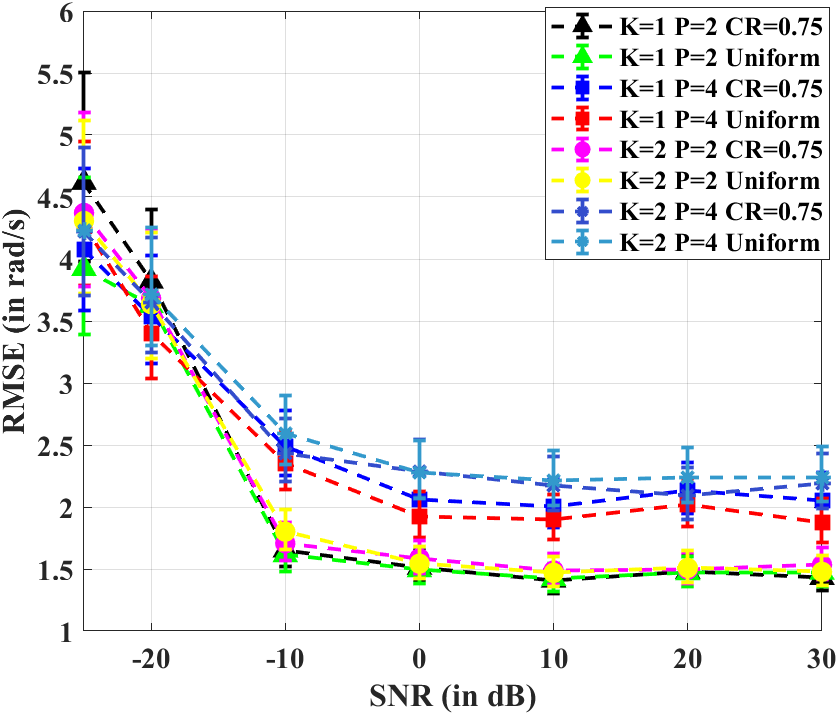}}
    \hfill
    \subfloat[]{\includegraphics[width=0.32\textwidth,height=4cm]{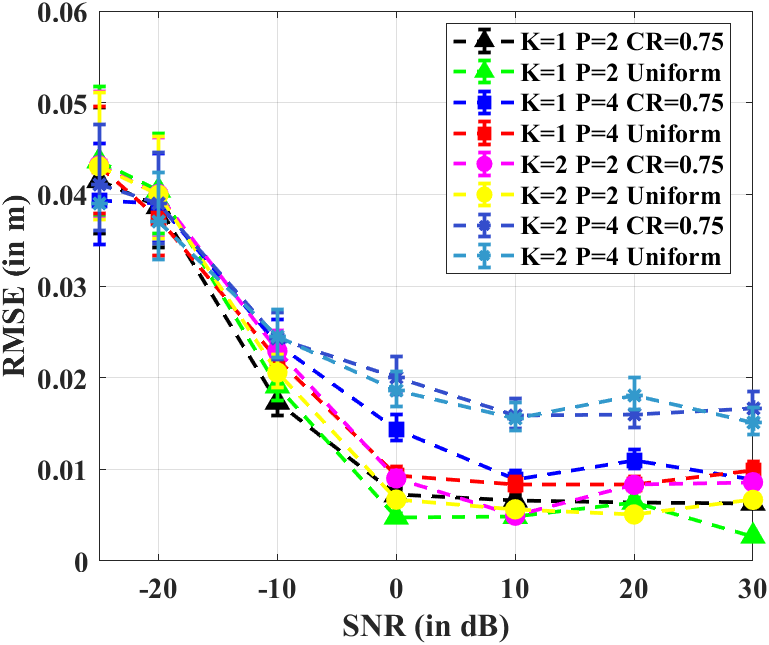}}
    \hfill
    \subfloat[]{\includegraphics[width=0.32\textwidth,height=4cm]{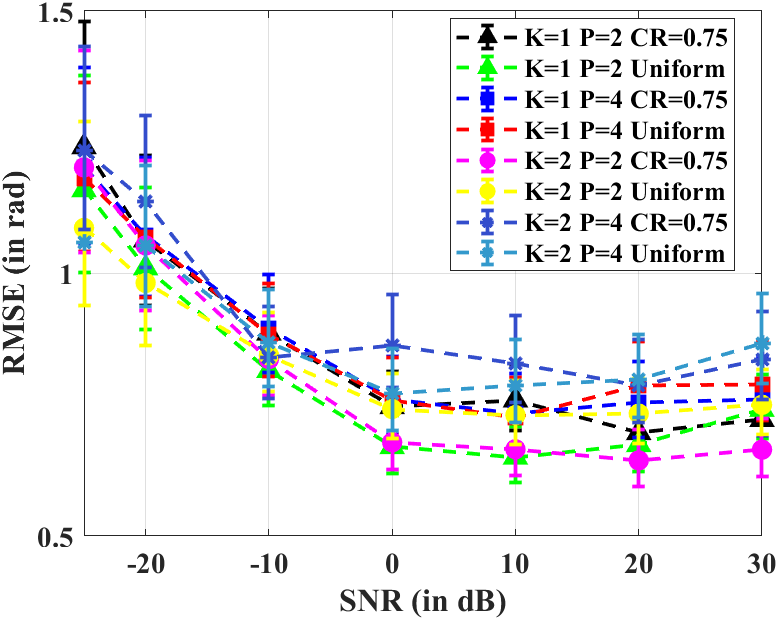}}
    \caption{RMSE of a) Rotational angular frequency ($\Omega$), b) Blade Length ($l_b$) and c) Initial phase angle ($\phi$) vs. SNR for~\nameref{Case I}. Error bars represent/denote the 95\% confidence interval.}
    \label{fig:rmse_K1P4}
\end{figure*}

\subsubsection{\textbf{Case I} } \textit{Targets Located in Separate Range Bins}\label{Case I} 

\begin{figure}[!t]
    \renewcommand{\thefigure}{6}
    \centering
    \includegraphics[width=0.75\linewidth]{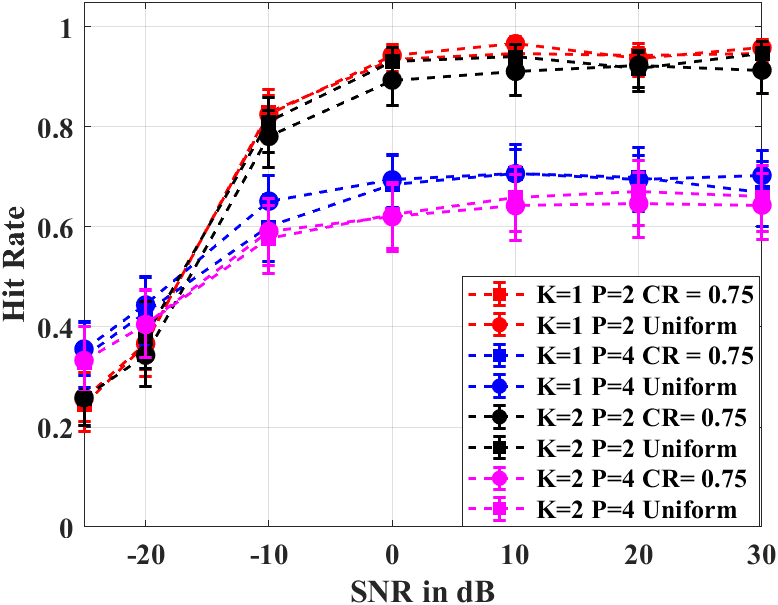}
    \caption{Comparison of Hit Rate vs SNR for CR=0.75 and uniform sampling in slow-time for a single target $(K=1)$ and multi-target $(K=2)$ scenarios, with $P=2$, $P=4$ for~\nameref{Case I}.}
    \label{fig:hit_false_rate_caseI}
\end{figure}

This section analyzes the estimation performance when target(s) occupy distinct range bins. Fig.~\ref{fig:hit_false_rate_caseI} show that, for both $K=1$ and $K=2$ with $P=2$, the proposed approach achieves a high hit rate of up to 0.96 in the uniform sampling case at higher SNR values.  The random sampling approach at $\text{CR = }0.75$ performs slightly better than the uniform sampling for $K=1$ with $P=2$ and gives comparable hit rate for $K=2$ with $P=2$. This can be attributed to the increased measurement incoherence provided by random sampling, which enhances the efficiency of compressive sensing-based recovery compared to uniform sampling along the slow-time domain. A consistent trend is observed as the target model increases in complexity, with 
$P$ varying from 2 to 4, as illustrated in Fig.~\ref{fig:hit_false_rate_caseI}. The hit rate gradually decreases with increasing $P$ because the sparsity of the underlying signal representation diminishes as additional scatterers or micro-motion components are introduced. Consequently, the estimation performance degrades in these more complex target conditions. However, since the RMSE is computed only for parameters that are correctly detected, its values remain relatively stable across different values of $K$, showing a deterioration only with increasing $P$ as shown in Fig.~\ref{fig:rmse_K1P4}. This stability demonstrates the algorithm’s robustness in estimating correctly identified parameters even when the overall hit rate decreases. The estimation accuracy of $l_b$ follows a similar trend as that of $\Omega$.

\begin{figure}[!t]
    \renewcommand{\thefigure}{8}
    \centering
    \includegraphics[width=0.75\linewidth]{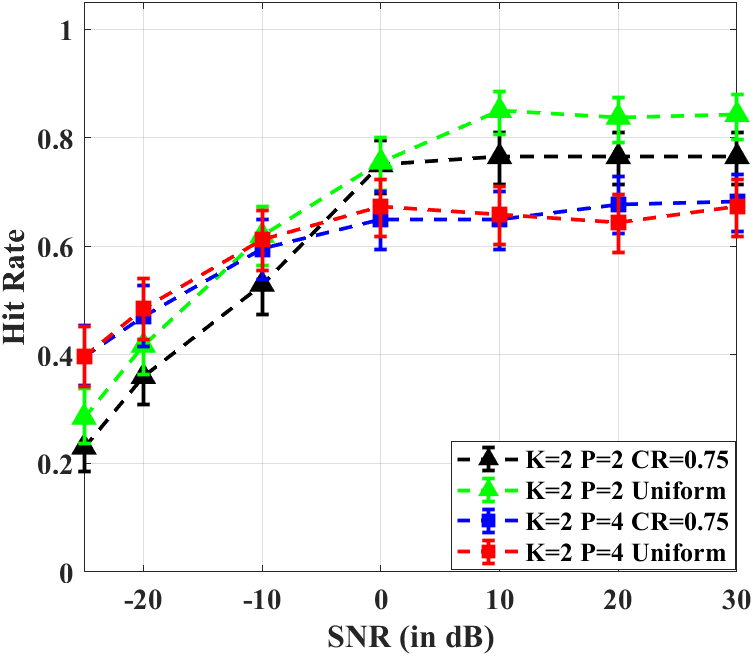}
    \caption{Comparison of Hit Rate vs SNR for CR=0.75 and uniform sampling in slow-time for a multi-target $(K=2)$ scenario, with $P=2$, $P=4$ for~\nameref{Case II}.}
    \label{fig:hit_false_rate_caseII}
\end{figure}

\begin{figure*}[!t]
    \renewcommand{\thefigure}{9}
    \centering
    \subfloat[]{\includegraphics[width=0.32\textwidth,height=4cm]{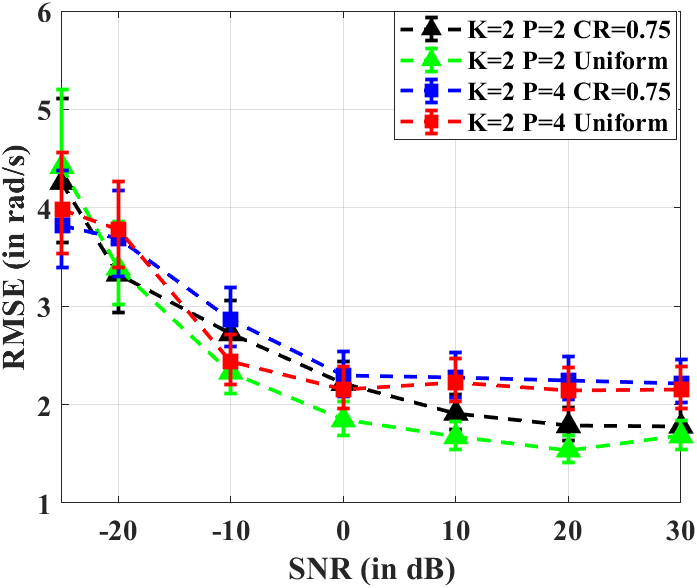}}
    \hfill
    \subfloat[]{\includegraphics[width=0.32\textwidth,height=4cm]{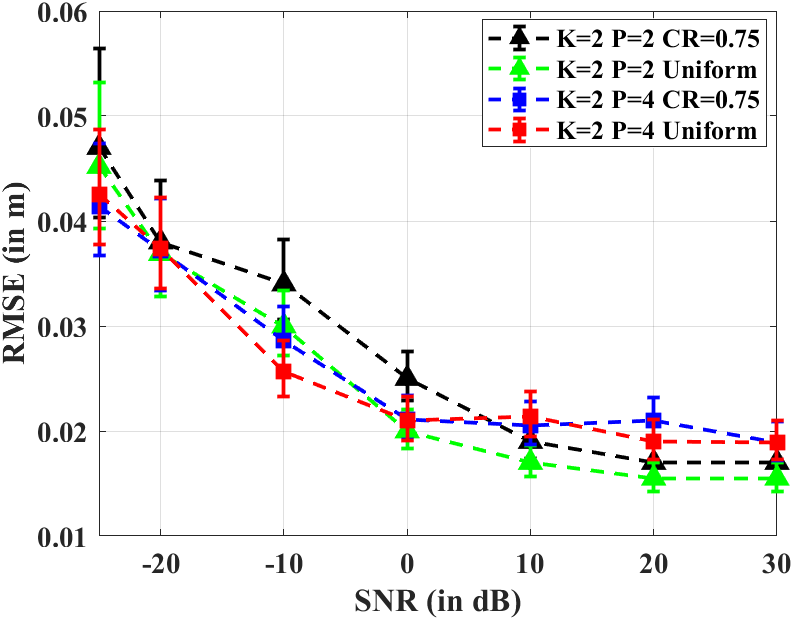}}
    \hfill
    \subfloat[]{\includegraphics[width=0.32\textwidth,height=4cm]{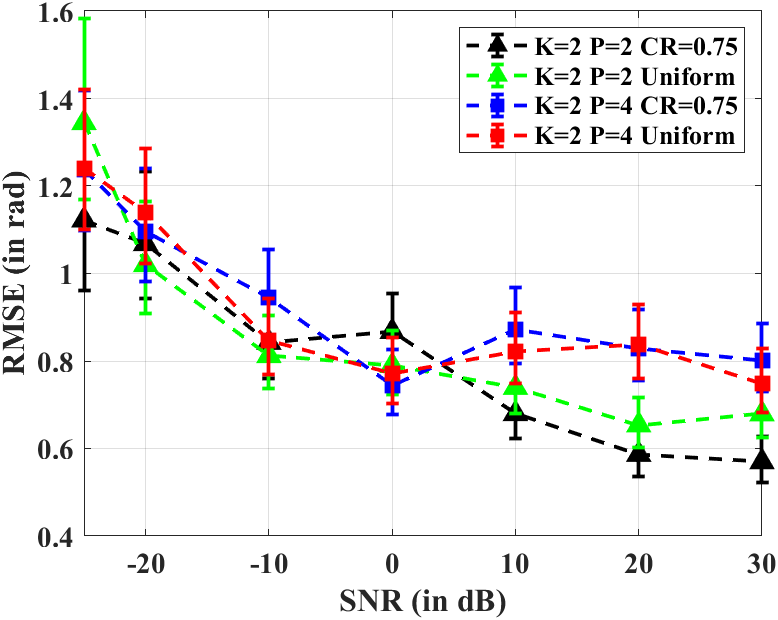}}
    \caption{a) RMSE of Rotational angular frequency ($\Omega$), b) RMSE of Blade Length ($l_b$) and c) Initial phase angle ($\phi$) vs. SNR for~\nameref{Case II}. Error bars represent/denote the 95\% confidence interval.}
    \label{fig:rmse_same}
\end{figure*}

\subsubsection{\textbf{Case II}} \textit{Multiple Targets in the Same Range Bin}
\label{Case II}

We now investigate the more challenging scenario in which multiple targets reside within the same range bin while possessing distinct Doppler, AoA, and micro-motion characteristics.

Unlike \nameref{Case I}, the co-located configuration considered here prevents effective suppression of the bulk component as both targets share the same estimated range frequency $\hat{f_r}$. This residual coupling between bulk and micro-motion components degrades the separability of the underlying micro-motion parameter sets. The effect of this incomplete bulk component suppression is most clearly reflected in the hit-rate performance as seen in Fig.~\ref{fig:hit_false_rate_caseII}. Across all simulated SNR conditions, the hit rate in \nameref{Case II} is consistently lower than the corresponding results for the target--propeller configurations in \nameref{Case I}, with no target configuration in \nameref{Case II} approaching the peak hit rate of $0.96$
observed in \nameref{Case I}. This degradation in estimation performance can be attributed to the presence of the strong bulk component, making it more difficult for the OMP-based estimator to extract the micro-Doppler signatures and correctly identify the individual parameter sets. A particularly noteworthy deviation from~\nameref{Case I} arises when comparing uniform and random sampling strategies. For the representative case of $K = 2$ and $P = 2$, the uniform sampling configuration outperforms the random-sampling case at $\text{CR} = 0.75$ in terms of the hit rate. This behavior contrasts with the findings in~\nameref{Case I}, where random sampling shows improved detection through enhanced measurement incoherence. A similar trend is observed in the RMSE performance as shown in Fig.~\ref{fig:rmse_same}. The uniform sampling configuration provides a lower RMSE than the random sampling case. However, the algorithm is capable of reliably detecting the targets and accurately estimating more than $60\,\%$ of the rotation frequencies with low RMSE, even in complex multi-target scenarios with reduced random measurements at SNR as low as $-10\,\mathrm{dB}$.

These observations collectively highlight the interplay between sparsity and measurement strategy in the proposed compressive sensing framework. At high SNR levels, sparse micro-Doppler features can be effectively recovered with limited measurements, leading to high hit rates and accurate parameter estimation. Random sampling proves advantageous by better capturing the incoherence required for successful sparse recovery. The benefits of random sampling are strongly dependent on the degree of sparsity and separability of the micro-Doppler signatures. While random sampling offers clear advantages when targets are distributed across different range bins, its performance degrades when multiple targets occupy the same range bin. In such cases, uniform sampling preserves the integrity of the measurements and yields a more reliable detection and parameter estimation performance. As the sparsity of the vector to be recovered decreases with increasing $P$, the recovery performance of the parametric dictionary deteriorates. This degradation can be attributed to the limitations of the dictionary in satisfying the Restricted Isometry Property (RIP), reducing its ability to distinguish between closely spaced signal components~\cite{GangLi2014}. \textcolor{red}. The simulations use $K = 2$ as a representative multi-target scenario, chosen for clarity while still capturing the key performance trends. The framework itself generalizes to $K > 2$; the dominant performance driver is the number of propeller blades $(P)$ rather than the number of targets $(K)$, since dictionary complexity scales with $P$. Additionally, by adopting an SLA configuration, the number of antenna elements is significantly reduced compared to a ULA with the same physical aperture, thereby improving system efficiency without compromising estimation accuracy~\cite{CSrai2025}. At high SNRs, the bulk parameter estimation performance of the proposed framework remains unaffected when employing a ULA as the spatial frequency estimation has been decoupled from the estimation of all other parameters. 
\begin{table}
\centering
\caption{Run Time Comparison}
\begin{tabular}{|c|c|c|c|}
\hline
\makecell{No. of Targets\\(\textbf{K})} & \makecell{No. of Propellers\\(\textbf{P})} & \textbf{CR = 0.75} & \textbf{Uniform} \\
\hline
1 & 2 & 1.940 s & 2.516 s \\
\hline
1 & 4 & 2.816 s & 3.668 s \\
\hline
2 & 2 & 4.186 s & 5.395 s \\
\hline
2 & 4 & 5.692 s & 7.383 s \\
\hline
\end{tabular}
\label{table:runtime}
\end{table}
 Also, Table~\ref{table:runtime} reports the algorithm runtime for both sampling strategies, showing a notable reduction for $\text{CR} = 0.75$. The above results are consistent with compressive sensing theory, which demonstrates that accurate recovery is attainable from fewer measurements under certain sparsity constraints.

\begin{figure*}[!t]
    \renewcommand{\thefigure}{10}
    \centering
    % --- Rotational Frequency (Omega) ---
    \subfloat[]{\includegraphics[width=0.24\textwidth]{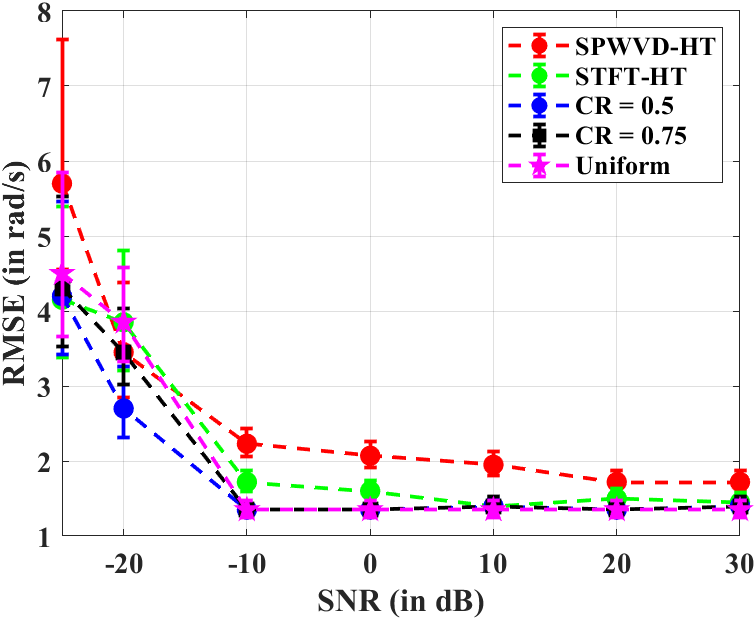}}
    \hfill
    \subfloat[]{\includegraphics[width=0.24\textwidth]{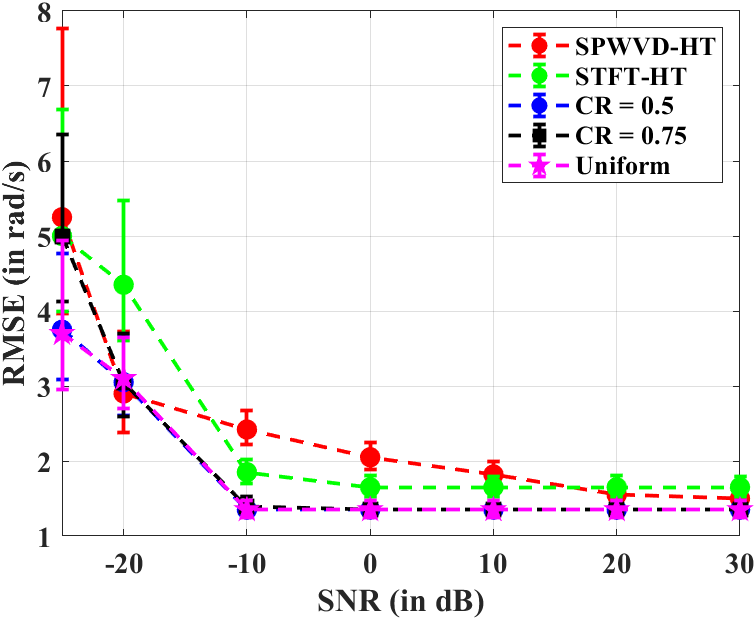}}
    \hfill
    % --- Blade Length (l_b) ---
    \subfloat[]{\includegraphics[width=0.24\textwidth]{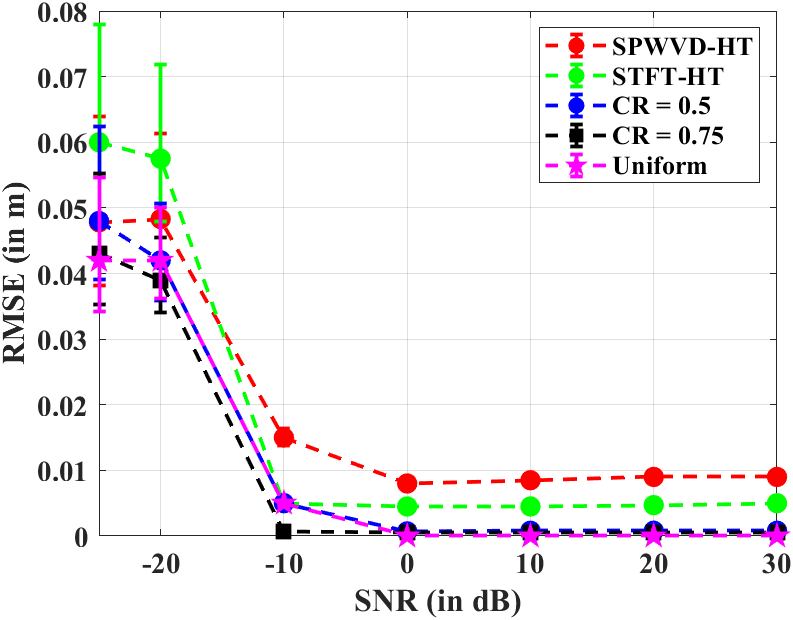}}
    \hfill
    \subfloat[]{\includegraphics[width=0.24\textwidth]{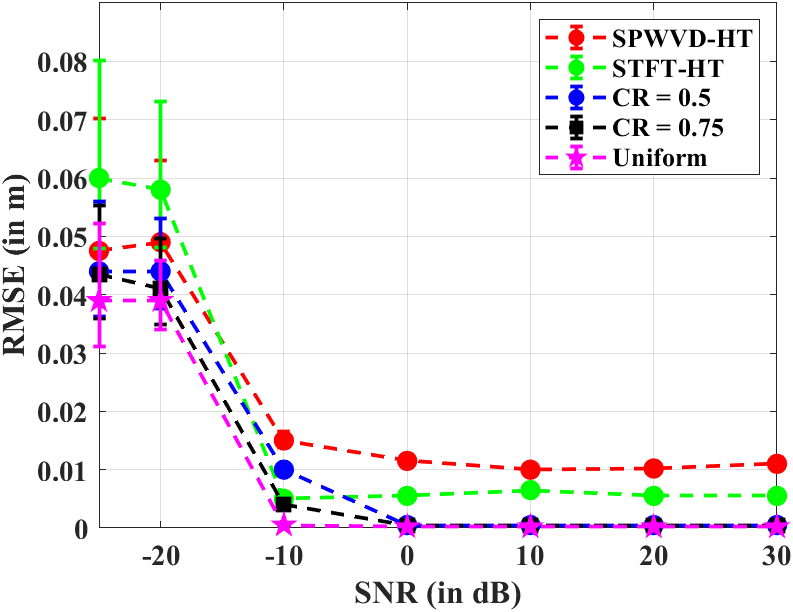}}
    \caption{RMSE of a) rotational frequency ($\Omega$) for $K=1$, $B=1$, b) rotational frequency ($\Omega$) for $K=1$, $B=2$, c) RMSE of blade length ($l_b$) for $K=1$, $B=1$ d) RMSE of blade length ($l_b$) for $K=1$, $B=2$ vs SNR using different methods. Error bars represent/denote the 95\% confidence interval.}
    \label{fig:tf_comparison}
\end{figure*}

\subsection{Comparison with TF methods}
\label{comparison tf}

\begin{figure}[!t]
    \centering
    \includegraphics[width=0.75\linewidth]{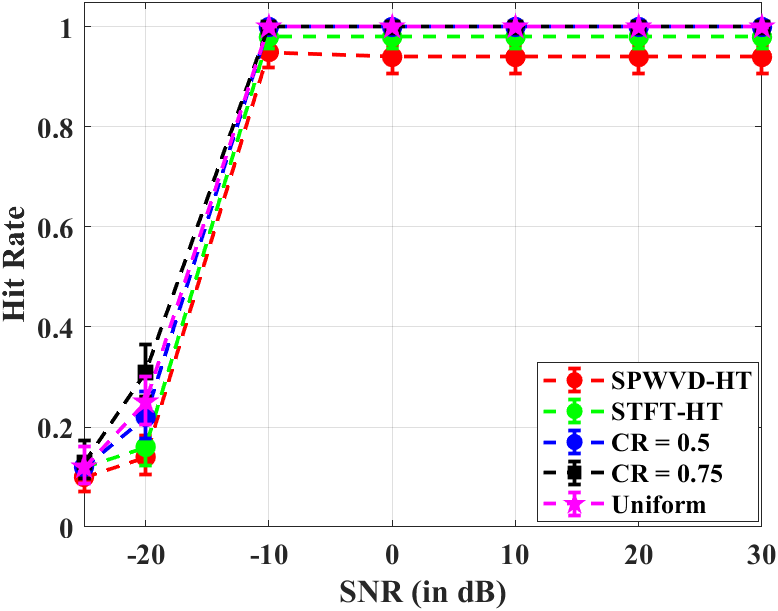}
    \caption{Comparison of Hit Rate vs SNR for $K=1$, $B=1$ using different methods.}
    \label{fig:hit_false_rate_TFI}
\end{figure}

\begin{figure}[!t]
    \centering
    \includegraphics[width=0.75\linewidth]{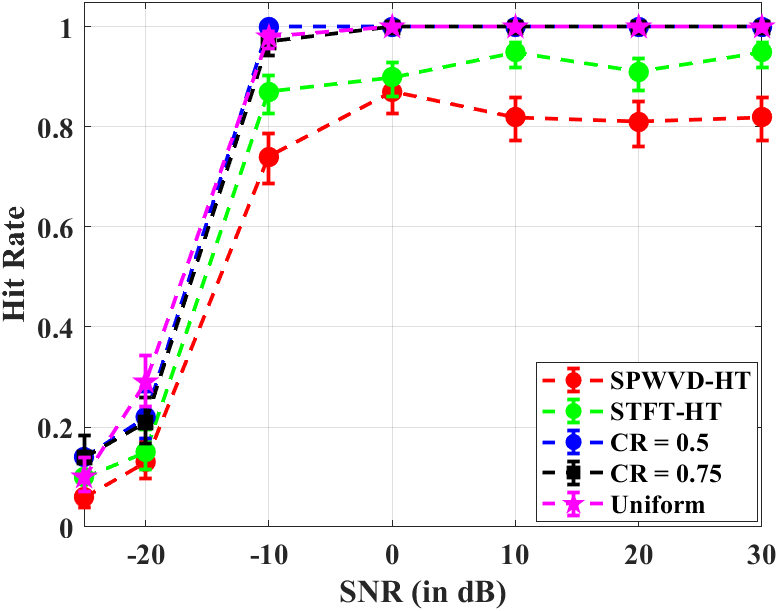}
    \caption{Comparison of Hit Rate vs SNR for $K=1$, $B=2$ using different methods.}
    \label{fig:hit_false_rate_TFII}
\end{figure}

We evaluate the parameter estimation accuracy of the proposed method against the TF-based algorithm presented in \cite{Barbarossa1996}, under the assumption that the optimal window length is known. Specifically, we employ the SPWVD and STFT as the underlying time–frequency representations, each followed by a Hough transform (HT) for parameter extraction. For clarity, these approaches are denoted as SPWVD–HT and STFT–HT, respectively, and are used to compare with the proposed framework. Here, the slow-time vector obtained after from step 8 in Algorithm~\ref{alg:algorithm1} is used as the input for these TF-based methods. This comparison is conducted under simplified settings involving a single target $(K=1)$, one propeller $(P=1)$, for one and two-rotating scatterers or blades $(B=1 \,\, \text{\&} \,\, B=2)$. Simulation results in Fig.~\ref{fig:tf_comparison}-~\ref{fig:hit_false_rate_TFII} demonstrate that under a CR of $0.5$, $0.75$ and uniform sampling, our method outperforms the TF-based methods in both the performance metrics i.e., hit rate and RMSE of both rotational frequency ($\Omega$) and radius of rotation ($l_b$). For the simplest case of $B=1$, the different methods have comparable hit rates with our algorithm performing slightly better. However, we observe that the TF-based methods struggle in complex scenarios involving multiple rotating scatterers $(B=2)$ because the accuracy of the Hough accumulation degrades when signal components are closely spaced in the TF domain. In contrast, our framework estimates the micro-motion parameters using a matching pursuit strategy, where each iteration selects a basis signal via correlation with the received signal followed by a least-squares projection. This enables coherent processing while suppressing interference from other components, resulting in improved estimation accuracy~\cite{GangLi2014}. However, a direct comparison with the other CS-based methods discussed in Section~\ref{Prior Art} is not feasible, as the underlying radar systems, corresponding signal models and dictionaries differ from those considered in this work.

\subsection{Classification Performance Evaluation}
% \vspace{-1em}
We divide the quadcopter movement into four distinct categories: Hover, Takeoff, Landing, and Translation (forward or backward motion). The estimated rotation frequencies \(\hat{\mathbf{\Omega}} = \left[ \hat{\Omega}_1 , \hat{\Omega}_2, \hat{\Omega}_3, \hat{\Omega}_4 \right]\) are first sorted in ascending order. The \textit{mean rotational frequency} and the \textit{front--rear frequency difference} are calculated from the second (representative of one set of propellers) and the third values (representative of the other set of propellers) of $\hat{\Omega}$ sorted in the ascending order, as these estimates are observed to be the most accurate for a target scenario with $P=4$. 

%A Monte Carlo simulation framework comprising $N=500$ trials per class was developed to evaluate classification robustness at an SNR of $10\text{dB}$. True propeller frequencies are generated according to quadcopter dynamics with class-specific signatures validated against small UAV propeller frequency and velocity specifications~\cite{Lehmann2022,dronespecs,Nasa2020}. The \textit{hover reference frequency} was set to $435\,\,\text{rad/s}$. The associated thresholds mentioned in Algorithm~\ref{alg:classifier} are given the following values: maximum frequency spread threshold $\Delta_{\Omega} = 10\,\text{rad/s}$, 
%hover deviation threshold $\Delta_{h} = 12\,\text{rad/s}$, 
%higher front--rear differential threshold $\Delta_{fr,h} = 25\,\text{rad/s}$, lower front--rear differential threshold $\Delta_{fr,l} = 12\,\text{rad/s}$, and velocity limits $v_{\text{low}} = 0.5\,\text{m/s}$ and $v_{\text{high}} = 6\,\text{m/s}$. 
%These parameters define permissible variations in propeller frequency and translational velocity under hover and translational motion conditions used during simulation. \textcolor{red}{Response 1E \& 4G: }\textcolor{blue}{The decision rules are established using the nominal rotor rotation frequencies, the changes in rotation frequency associated with hover and translational flight, and the characteristic ascent and descent velocities of small commercially available UAVs. The references supporting these parameter ranges have been included.}

A Monte Carlo simulation framework comprising $N=500$ trials per class is developed to evaluate classification robustness at an SNR of $10\,\text{dB}$. True propeller frequencies are generated according to quadcopter dynamics with class-specific signatures, using a hover reference frequency of $435\,\text{rad/s}$, consistent with reported rotor speed ranges for small commercial UAVs~\cite{Lehmann2022}. The velocity limit $v_{\text{high}} = 6\,\text{m/s}$ corresponds to the nominal ascent speed reported for small commercial UAV platforms~\cite{dronespecs}, while $v_{\text{low}} = 0.5\,\text{m/s}$ is chosen to distinguish hover-induced jitter from intentional translational motion. The remaining decision thresholds in Algorithm~\ref{alg:classifier}, namely the maximum frequency spread $\Delta_{\Omega} = 10\,\text{rad/s}$, hover deviation $\Delta_{h} = 12\,\text{rad/s}$, higher front--rear differential $\Delta_{fr,h} = 25\,\text{rad/s}$, and lower front--rear differential $\Delta_{fr,l} = 12\,\text{rad/s}$, are selected as small fractions (approximately $2$--$8\%$) of the full rotor operating range observed across small commercial UAVs~\cite{Lehmann2022,Nasa2020}, and are further tuned through simulation to ensure reliable discrimination between hover and translational flight regimes.

\begin{table}
\centering
\caption{Confusion Matrix (\%)}
\label{tab:confusion}
\renewcommand{\arraystretch}{1.3} % Increase row height
\setlength{\tabcolsep}{5pt} % Adjust column spacing
\fontsize{9}{10}\selectfont % Increase table font slightly (9pt)
\begin{tabular}{lcccc}
\hline
\multirow{2}{*}{\textbf{True$\backslash$Pred}} & \multicolumn{4}{c}{\textbf{Predicted Class}} \\
\cline{2-5}
 & Hover & Landing & Takeoff  & Translation \\
\hline
Hover     & \textbf{97.22} & 0.39 & 0 & 2.38 \\
Landing   & 3.12 & \textbf{95.2} & 0 & 1.66 \\
Takeoff   & 2.42 & 0 & \textbf{96.76} & 0.8 \\
Translation & 0 & 0 & 1.5 & \textbf{98.5} \\
\hline
\end{tabular}
\label{tab:classfication}
\end{table}

\begin{figure}[!t]
    \centering
    \includegraphics[width=\linewidth]{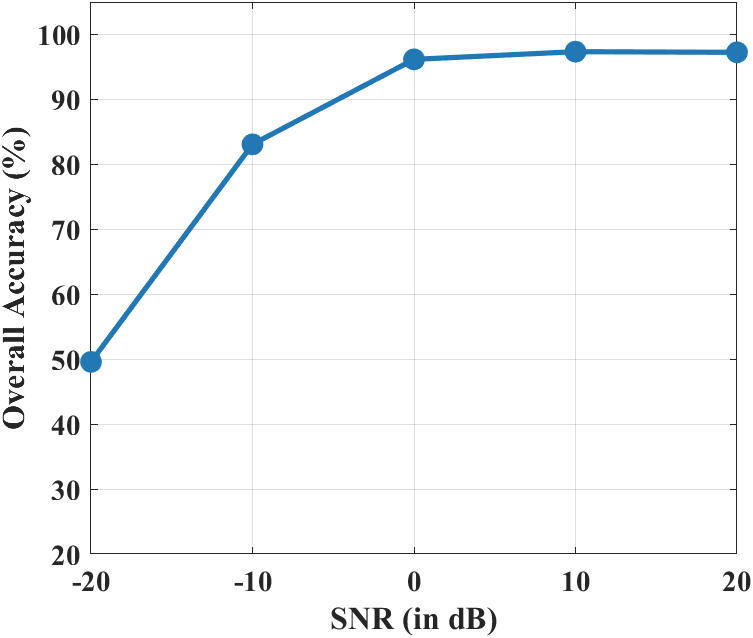}
    \caption{Overall classification accuracy vs SNR.}
    \label{fig:accuracy vs SNR}
\end{figure}

Table~\ref{tab:classfication} presents the row-normalized confusion matrix obtained using the proposed rule-based classifier (Algorithm~\ref{alg:classifier}). The framework achieves an overall classification accuracy of 97\%. The \textit{Translation} mode exhibits the highest detection accuracy (98.5\%), attributed to the distinct front–rear rotation frequency differential and velocity-based decision rule incorporated in the algorithm. \textit{Hover}, \textit{Takeoff}, and \textit{Landing} classes also demonstrate high accuracies above 95\%. Most misclassifications occur between the \textit{Hover} and vertical motion classes due to their similar rotor dynamics and body motion characteristics. The overall classification accuracy with respect to different SNR conditions are presented in Fig~\ref{fig:accuracy vs SNR}. These results highlight the effectiveness of the proposed MIMO–FMCW-based micro-motion parameter estimation framework for reliable UAV flight mode identification. Nevertheless, capturing more complex and nonlinear quadcopter maneuvers will require an extended and physically grounded signal modeling framework to improve generalization.

\FloatBarrier

\section{Conclusion}
\label{conclusion}
This work demonstrates that accurate micro-motion parameter estimation for small quadcopters can be achieved using a sparsity-driven MIMO-FMCW radar framework with significantly reduced measurement requirements. By exploiting sparsity across the range, Doppler, spatial, and micro-Doppler frequency domains, the proposed approach reliably estimates key parameters including range, radial velocity, AoA, rotational frequency and blade length. The results show that precise micro-motion estimation is possible using a randomly selected subset of slow-time chirps transmitted from a MIMO SLA, without relying on time–frequency transforms that require uniformly spaced samples. Furthermore, the improved AoA estimation enabled by the MIMO-FMCW configuration enhances the accuracy of blade length inference, as the Doppler modulations are inherently AoA dependent. Finally, the estimated micro-motion parameters and the radial velocity are used to classify the motion type of the UAV effectively using a set of decision rules.

Future efforts will focus on implementing the proposed algorithm in hardware and validating its performance using real radar measurements acquired from quadcopter platforms. Subsequent work will also incorporate a more physically accurate, orientation-dependent RCS model that captures the amplitude modulation introduced as each blade rotates relative to the radar line of sight.

\section*{Appendix A}
\label{appendixa}
\subsection{Derivation of the IF Signal with Simplifying Approximations}
Let us consider a case where there is a single target with a single blade, allowing us to disregard the index $k$ and $b$. Mixing the received signal in \eqref{eqn:received_signal} with the transmitted waveform in \eqref{eqn:transmit_signal} yields the intermediate-frequency (IF) signal corresponding to a single blade, which is given by,
\begin{align}
q_{\textit{rx},\textit{tx},l}^{b}(t) &= a^*\cdot s^*(t-(\tau_{\textit{rx},\textit{tx},l}^R + \tau_{\textit{rx},\textit{tx},l}^\theta))\cdot s(t) \notag\\
 &= e^{-j2\pi f_c(t-(\tau_{\textit{rx},\textit{tx},l}^R + \tau_{\textit{rx},\textit{tx},l}^\theta))}\cdot e^{-j \pi\gamma(t-(\tau_{\textit{rx},\textit{tx},l}^R + \tau_{\textit{rx},\textit{tx},l}^\theta))^2} \notag \\ &\quad\times e^{j2\pi f_c t} \cdot e^{j\pi \gamma t^2}\notag \\
 &= \underbrace{e^{j2\pi f_c(\tau_{\textit{rx},\textit{tx},l}^R + \tau_{\textit{rx},\textit{tx},l}^\theta)}}_{\text{Term I}} \cdot \underbrace{e^{j\pi2\gamma(\tau_{\textit{rx},\textit{tx},l}^R + \tau_{\textit{rx},\textit{tx},l}^\theta)t)}}_{\text{Term II}} \notag \\ &\quad\times \underbrace{e^{-j\pi\gamma(\tau_{\textit{rx},\textit{tx},l}^R + \tau_{\textit{rx},\textit{tx},l}^\theta)^2}}_{\text{Term III}}\label{eqn:appendix}
\end{align}
We analyze each term of \eqref{eqn:appendix} after expanding it using \eqref{eqn:roundtrip_range} and \eqref{eqn:roundtrip_space}. The values of the parameters used for this analysis are $R_0 = 120\,m$, $v = 70\,m/s$, $L= 256$, $T_c = 40 \,\mu s$, $T_s = 2\times10^{-7}\,s$, $\gamma = 6.25\times10^{12} \,s^{-2}$, $Z = 6\lambda\,m$, $l_b = 0.15\,m$ and $B = 250\times10^{6} Hz$.
\begin{enumerate}
    \item \textbf{Term I}: This term can be further expanded as 
    \begin{align}
    e^{j2{\pi} {f_c}\tau_{\textit{rx},\textit{tx},l}(t)} = e^{j 2 \pi f_c \frac{2 R_0}{c}} \cdot e^{j 2 \pi f_c \frac{2vl\cdot T_c}{c}} \\ \notag 
    \times e^{j 2 \pi f_c \frac{2l_b\cos(\theta) \cos(\Omega(t))}{c}}\cdot
    e^{j 2 \pi f_c \frac{ Z(\alpha_\textit{tx} + \beta_\textit{rx})\sin\theta}{2c}}.
    \end{align}
    The term $e^{j 2 \pi f_c \frac{2 R_0}{c}}$ is absorbed into the constant $a^*$ as it is independent of time. The phase terms of $e^{j 2 \pi f_c \frac{2vl \cdot T_c}{c}}$ , $e^{j 2 \pi f_c \frac{2l_b\cos(\theta) \cos(\Omega(t))}{c}}$ and $e^{j 2 \pi f_c \frac{ Z(\alpha_\textit{tx} + \beta_\textit{rx})\sin\theta}{2c}}$ are of significant order and  time-dependent. Therefore, it is considered for further analysis. \\
     \item \textbf{Term II}: This term can be expanded as
     \begin{align}
     \label{eqn:term2}
      e^{j2{\pi} \gamma\tau_l(t)\cdot t } &= e^{j 2 \pi \gamma  \frac{2 R_0}{c}t} \cdot e^{j 2 \pi \gamma \frac{2vl\cdot T_c}{c}t}\\ \notag 
      &\times e^{j 2 \pi \gamma \frac{2l_b\cos(\theta) \cos(\Omega(t))}{c}t}\cdot e^{j 2 \pi \gamma \frac{ Z(\alpha_\textit{tx} + \beta_\textit{rx})\sin\theta}{c}}
      \end{align}
    It can be observed that the phase value of the term \( e^{j 2 \pi \gamma  \frac{2 R_0}{c}t} \) is approximately:
    \[
    2\pi \cdot 6.25 \times 10^{12} \cdot 1.6 \cdot 10^{-13} \approx 2\pi ,
    \]
    and this phase increases with the number of fast-time samples.
    Similarly, the phase value of the term \( e^{j 2 \pi \gamma \frac{2vl \cdot T_c}{c}t} \) is
    \[
    2\pi \cdot 6.25 \times 10^{12} \cdot 2\cdot70\cdot40\cdot2 \times 10^{-7} /3\cdot10^{8} \approx 1.46\times10^{-4} .
    \]
    For the term \( e^{j 2 \pi \gamma \frac{2l_b \cos(\theta)  \cos(\Omega(t))}{c}t} \), the phase value is,
    \[
    2\pi \cdot 6.25 \times 10^{12}\cdot 2 \cdot0.15\cdot 2\cdot 10^{-7}/3\cdot10^{8}  \approx 7.84\times10^{-3}.
    \]
     For the term \( e^{j 2 \pi \gamma \frac{ Z(\alpha_\textit{tx} + \beta_\textit{rx})\sin\theta}{c}}\), the phase value is,
    \[
    2\pi \cdot 6.25 \times 10^{12}\cdot 6\cdot 0.0125\cdot 2\cdot 10^{-7}/2\cdot3\cdot10^{8}  \approx 9.81\times10^{-4}.
    \]
    
    Therefore, from \eqref{eqn:term2}, $e^{j 2 \pi \gamma  \frac{2 R_0}{c}t}$ and $e^{j 2 \pi \gamma\frac{2l_b\cos(\theta) cos(\Omega(t))}{c}t}$ are included for further analysis. Although the term that includes micro-motion has a much lower phase value compared to the range beat frequency, we keep it in the analysis to see the effect of micro-motion.
    
    \item \textbf{Term III}: We can expand \(e^{-j{\pi} \gamma \tau_{\textit{rx},\textit{tx},l}(t)^2}\) as \(e^{-j{\pi} \gamma ((\tau_{\textit{rx},\textit{tx},l}^{R})^2 + (\tau_{\textit{rx},\textit{tx},l}^{\theta})^2 + 2\cdot\tau_{\textit{rx},\textit{tx},l}^{R}\cdot \tau_{\textit{rx},\textit{tx},l}^{\theta})} \). The phase term is, however, dominated by the square of the range term, and this makes  $(\tau_{\textit{rx},\textit{tx},l}^{R})^2 >> (\tau_{\textit{rx},\textit{tx},l}^{\theta})^2$. Within $(\tau_{\textit{rx},\textit{tx},l}^{R})^2$, time independent term $R_0 ^2$ dominates the phase values and is absorbed into $a^*$~\cite{CSrai2025}. The rest of the terms in term III can be neglected for further analysis.\\
    
\end{enumerate}
Hence, the effective IF signal is given by, 
\begin{align}
    q_{\textit{rx},\textit{tx},l}(t) &= a^{*} \cdot e^{j 2 \pi f_c \frac{Z (\alpha_\textit{tx} + \beta_\textit{rx}) \sin(\theta)}{2c}} \label{eqn:blade_signal_approx}
 \\
&\quad \times e^{j 2 \pi \gamma \frac{2 R_0}{c} t} 
\cdot e^{j 2 \pi \gamma \frac{2 l_b \cos\theta \cos\left(\Omega_p(t)\right)}{c} t} \notag \\
&\quad \times e^{j 2 \pi f_c \frac{2 v}{c} t} 
\cdot e^{j 2 \pi f_c \frac{2 l_b \cos\theta \cos\left(\Omega_p(t) \right)}{c}}. \notag    
\end{align}

\bibliographystyle{IEEEtran}
\bibliography{references}

\end{document}